\documentclass[12pt]{SelfArx} 
\usepackage[english]{babel} 
\usepackage{listings}
\usepackage{lscape}

\definecolor{color1}{RGB}{0,0,90} 
\definecolor{color2}{RGB}{0,20,20} 

\usepackage[dvipdfm]{hyperref} 
\hypersetup{hidelinks,colorlinks,breaklinks=true,urlcolor=red,citecolor=color1,linkcolor=color1,bookmarksopen=false,
	pdftitle={An efficient and robust method for analyzing population pharmacokinetic data in genome-wide pharmacogenomic studies: a generalized estimating equation approach},
	pdfauthor={Kengo Nagashima, Yasunori Sato, Hisashi Noma, Chikuma Hamada}}

\JournalInfo{\textit{Statistics in Medicine} 2013;\textbf{32}(27):4838--4858.} 
\Archive{DOI: 10.1002/sim.5895} 

\PaperTitle{An efficient and robust method for analyzing population pharmacokinetic data in genome-wide pharmacogenomic studies: a generalized estimating equation approach}
\ShortTitle{An efficient and robust method for analyzing population pharmacokinetic data}

\Authors{Kengo Nagashima\textsuperscript{1,2,3}*, Yasunori Sato\textsuperscript{1,4}, Hisashi Noma\textsuperscript{5}, and Chikuma Hamada\textsuperscript{6}} 
\affiliation{\footnotesize{\textsuperscript{1}\textit{Clinical Research Center, Chiba University Hospital, 1-8-1 Inohana, Chuo-ku, Chiba 260-8677, Japan.}}}
\affiliation{\footnotesize{\textsuperscript{2}\textit{Graduate School of Engineering, Tokyo University of Science, 1-3 Kagurazaka, Shinjuku-ku, Tokyo 162-8601, Japan.}}}
\affiliation{\footnotesize{\textsuperscript{3}\textit{Faculty of Pharmaceutical Sciences, Josai University, 1-1 Keyakidai, Sakado-shi, Saitama 350-0295, Japan.}}}
\affiliation{\footnotesize{\textsuperscript{4}\textit{Department of Biostatistics, Harvard School of Public Health, Boston, Massachusetts 02115, U.S.A.}}}
\affiliation{\footnotesize{\textsuperscript{5}\textit{Department of Data Science, The Institute of Statistical Mathematics, 10-3 Midori-cho, Tachikawa, Tokyo 190-8562, Japan.}}}
\affiliation{\footnotesize{\textsuperscript{6}\textit{Faculty of Engineering, Tokyo University of Science, 1-3 Kagurazaka, Shinjuku-ku, Tokyo 162-8601, Japan.}}}
\affiliation{\footnotesize{*\textbf{Corresponding author}: nshi1201@gmail.com}}
\affiliation{}
\affiliation{This is the postprint / accepted version (24-May-2013) of the following article: ``Nagashima K, Sato Y, Noma H, Hamada C. An efficient and robust method for analyzing population pharmacokinetic data in genome-wide pharmacogenomic studies: a generalized estimating equation approach. \textit{Statistics in Medicine} 2013;\textbf{32}(27):4838--4858. DOI: 10.1002/sim.5895'', which has been published in final form at: \href{https://doi.org/10.1002/sim.5895}{https://doi.org/10.1002/sim.5895}. This article may be used for non-commercial purposes in accordance with Wiley Terms and Conditions for Self-Archiving.}

\Keywords{
\footnotesize{
Gene screening;
Generalized Estimating Equations;
Genome-wide study;
Misspecified models;
Population pharmacokinetic data.
}
} 
\newcommand{\keywordname}{Keywords} 

\Abstract{
\footnotesize{
Powerful array-based single-nucleotide polymorphism--typing platforms have recently heralded a new era in which genome-wide studies are conducted with increasing frequency.
A genetic polymorphism associated with population pharmacokinetics (PK) is typically analyzed using nonlinear mixed-effect models (NLMM).
Applying NLMM to large-scale data, such as those generated by genome-wide studies, raises several issues related to the assumption of random effects, as follows:
(i) Computation time: it takes a long time to compute the marginal likelihood.
(ii) Convergence of iterative calculation: an adaptive Gauss--Hermite quadrature is generally used to estimate NLMM; however, iterative calculations may not converge in complex models.
(iii) Random-effects misspecification leads to slightly inflated type-I error rates.
As an alternative effective approach to resolving these issues, in this article we propose a generalized estimating equation (GEE) approach for analyzing population PK data.
In general, GEE analysis does not account for inter-individual variability in PK parameters; therefore, the usual GEE estimators cannot be interpreted straightforwardly, and their validities have not been justified.
Here, we propose valid inference methods for using GEE even under conditions of inter-individual variability, and provide theoretical justifications of the proposed GEE estimators for population PK data.
In numerical evaluations by simulations, the proposed GEE approach exhibited high computational speed and stability relative to the NLMM approach.
Furthermore, the NLMM analysis was sensitive to the misspecification of the random-effects distribution, and the proposed GEE inference is valid for any distributional form.
An illustration is provided using data from a genome-wide pharmacogenomic study of an anticancer drug.
}
}

\begin{document}

\flushbottom 

\maketitle 


\thispagestyle{empty} 


\section{Introduction}

Individual variations in drug efficacy and side effects pose serious problems in medicine.
These variations are influenced by factors such as drug-metabolizing enzymes, drug transporters, and drug targets (e.g., receptors).
For many medications, these factors can be attributed to genetic polymorphisms \cite{Evans2001,Evans2003}.
Indeed, these genomic biomarkers are sometimes used to improve drug responses and reduce side effects by controlling the medication or dose according to the patient's genotype \cite{Innocenti,Wilkinson}.

However, only a few of these genomic biomarkers have been validated.
For this reason, many pharmacogenomics (PGx) studies have been launched around the world.
The purpose of these studies is to identify genes that affect drug-metabolizing enzymes, drug transporters, and drug targets.
Therefore, pharmacokinetics (PK) studies that include analyses of single-nucleotide polymorphisms (SNPs) as genomic markers in candidate-gene or genome-wide studies can be used to identify these genes.
The availability of powerful array-based SNP-typing platforms has facilitated genome-wide studies, which have become a standard strategy.
Such platforms make available to researchers genotype data for 100,000--4,300,000 SNPs.

In PK studies, it is common to apply compartmental models, which are often nonlinear models that include several PK parameters, in order to describe the profiles of drug concentrations in blood \cite{Wagner}.
Because drug concentrations in blood are usually related to drug efficacy and side effects, via their interactions with drug-metabolizing enzymes, drug transporters, and drug targets, differences in PK parameters indicate differences in effectiveness and toxicity.
Therefore, one object of PGx studies is to identify genes associated with PK parameters.

Because drug-concentration data is measured from multiple subjects in PGx studies, inter-individual variability in PK parameters should be considered.
Such data is referred to as population PK data.
If the impact of inter-individual variability in model parameters is ignored, no statistically valid inference is possible.
The mixed-effects model, which includes both fixed and random effects, is one method that accounts for inter-individual variability.
Inter-individual variability in model parameters is modeled as random effects with strong parametric assumptions about the random-effects distribution.
These models have often been used in analysis of longitudinal data \cite{Laird}, and they represent a useful method for accounting for inter-individual variability.
Moreover, the nonlinear mixed-effects model (NLMM), an extension of the mixed-effects model to nonlinear functions, is often used to analyze population PK data \cite{Lindstrom,Davidian1993,Davidian1995,Vonesh,Wolfinger}.

The association between PK parameters and SNPs are typically analyzed using an NLMM \cite{Hesselink,Bosch,Bertrand2009,Bertrand2012} in conjunction with population PK data.
However, applying an NLMM to large-scale data can be problematic for the following reasons:
\begin{enumerate}
	\item[(i)] Computation time:
	NLMMs can be computationally intensive, because these models must compute the marginal log-likelihood by integrating out random effects \cite{Davidian1993}.
	In NLMMs, inferences about model parameters are based on the marginal log-likelihood function, which includes a multiple integral with respect to the unobservable random effects.
	Because the regression functions are non-linear, the integral in the marginal log-likelihood function has no closed form, and it is necessary to compute the integral.
	To address this issue, various methods have been proposed to compute the integral approximation.
	However, these methods are computationally inefficient.
	\item[(ii)] Convergence of iterative calculations:
	For instance, a major statistical software package, the SAS NLMIXED Procedure (SAS Institute, Inc., Cary, North Carolina) with adaptive Gauss--Hermite quadrature, is now used to approximate the maximum marginal log-likelihood \cite{Pinheiro,Zhang}.
	These computations are based on iterative calculations; for complex models, however, these calculations may not converge \cite{Hartford,Lesaffre}.
	If iterative calculations do not converge, we derive no information from valuable data.
	\item[(iii)] Random-effects misspecification:
	Random-effects misspecification leads to bias in parameter estimates of the regression coefficients, and slightly inflates type-I error rates of tests for the regression coefficients in generalized mixed-effects models \cite{Neuhaus,Heagerty,Litiere2007,Litiere2008} and NLMMs \cite{Hartford}.
	Therefore, careful model building and checking are needed for each of the 100,000--4,300,000 analyses, but in practice this may be difficult to apply.
\end{enumerate}
In conclusion, it seems that these three problems occur in association with a strong assumption of random effects.

Therefore, we consider a new method that uses a potentially misspecified model to avoid the strong assumptions of the random-effects distribution.
Misspecified models are useful and powerful tools for studying the behavior of estimators under model misspecification.
Model misspecification means that an incorrect working model is used for estimation.
In this paper, we consider a ``true'' model that includes inter-individual variability in model parameters as fixed-effect parameter vectors $\boldsymbol{\beta}_i$ for each subject $i$ ($=1, 2, \ldots, K$), and a ``working'' model that misspecifies the presence of inter-individual variability in model parameters as a common parameter vector $\boldsymbol{\beta}$.
In this paper, we describe a new interpretation of the estimator $\widehat{\boldsymbol{\beta}}$ as a weighted average of the individual parameter vectors $\boldsymbol{\beta}_i$.
The proposed method allows for computation that is faster than NLMM by a factor of 100, performs stable computations, and is robust for various structures of individual variations, because it is based on misspecified fixed-effect models instead of random-effect models.

A general theory of misspecified models has been proposed by White \cite{White} for maximum-likelihood methods, and by Yi and Reid \cite{Yi} for estimating equations.
In both papers, the authors demonstrate the asymptotic normality of estimators of working-model parameters under mild conditions.
The method we propose uses generalized estimating equations (GEE) of a working model based on Yi and Reid's result.
GEE has been widely used in regression analyses of the generalized linear models with correlated response, such as repeated-measurement data \cite{Liang,Zeger1986,Aerts}.
Under mild conditions, the estimator from GEE of a misspecified model is consistent and asymptotically normal.
Therefore, the proposed method may be applied to correlated-response data with inter-individual variability in model parameters, which includes a wide range of applications.
In this paper, however, we focus on the problem of estimating PK parameters in the presence of inter-individual variability in PGx studies, i.e., the motivating example.

The proposed method focuses only on estimating fixed effect parameter vectors, because one object of PGx studies is to identify genes associated with PK parameters.
Other parameters are nuisance parameters.
Therefore, a misspecified model that gives an estimator for a weighted average of fixed effect parameter vectors is ``intentionally'' used.
As a result, the proposed estimator $\hat{\boldsymbol{\beta}}$ is different from the estimator of NLMMs.
Marginal (or population-averaged) models and mixed (or subject-specific) models can lead to a different estimator in non-linear settings \cite{Zeger1988}.
The proposed method relates to marginal models.

For each SNP, there are three genotypes for each locus: the ``aa'', ``Aa'', and ``AA'' genotypes, where ``a'' is the major allele and ``A'' is the minor allele.
There are often a considerable number of genes for which the frequency of the minor homozygous genotype, ``AA'', is very small, because a SNP is defined as a mutation involving a single DNA base substitution that is observed with a frequency of at least 1{\%} in a population.
Therefore, for valid statistical inference, a small--sample correction is needed.
To address the small--sample size problem, we propose a Wald-type test and an asymptotic $F$-test for determining the effects of a genetic polymorphism on PK parameters (see Section 3.5).

In Section 2, motivating data and issues of NLMM are introduced.
In Section 3, misspecified models and the proposed method are presented, and some of the proposed method's theoretical properties are discussed.
In Section 4, we study the performance of the proposed method using simulations.
In Section 5, we present the application of the method to published experimental data.
We present our concluding remarks in Section 6.

\section{Motivating example}

The motivation for this paper stems from an PGx study data \cite{Sato} on gemcitabine (2',2'-difluorodeoxy\-cytidine), which is a nucleoside anticancer drug.
The study was designed to screen for genes related to the PK of gemcitabine.
The participants consisted of 233 gemcitabine-naive cancer patients (mainly with pancreatic carcinoma).
For the PK analysis, heparinized blood samples were taken before administration and at 0.5, 0.75, 1.0, 1.5, 2.0, 2.5, and 4.5 hours after the beginning of the administration.
The dose was adjusted according to the surface area of the body of each subject.
A total of 109,365 gene-centric SNPs were genotyped using the Sentrix Human-1 Genotyping BeadChip (Illumina Inc., San Diego, CA).

Because the main object of this PGx study was to screen PK-related genes, the SNP genotype effect on PK parameters was modeled using a compartmental model.
Compartmental models, which are derived from differential equations that describe drug kinetics, are nonlinear models with several PK parameters.
It is common to apply such models in order to describe the profiles of drug concentrations in blood.

In general, analyses of genome-wide data use appropriate statistical methods to investigate the association between an outcome variable and a set of SNPs \cite{Ziegler}.
Based on the results, favorable SNPs that strongly associate with the outcome variable are screened with appropriate criteria (e.g., the Bonferroni adjustment, false-discovery rate, etc.).
The appropriate statistical methods are determined by the nature of the outcome variable and the study design (e.g., trend tests for odds ratios are used in case-control studies), and these analyses are commonly performed one by one for each SNP.
In genome-wide PGx studies, to identify genes that associate with PK parameters, analyses of the associations between PK parameters and SNPs are applied to population PK data.
We consider that the SNP genotype effect reflects the difference in average PK parameters between different genotypes.
Moreover, because the PK data include multiple individuals, the data are population PK data, and we must consider the impact of inter-individual variability in PK parameters.

Now, we introduce notations and describe the data structure.
Suppose we have a plasma drug concentration dataset with $K$ subjects and a genotype dataset with $M$ SNPs.
For each subject $i$ ($=1, 2, \ldots, K$), there is a random ($n_i \times 1$)-dimensional vector $\mathbf{Y}_i=(Y_{i1}, Y_{i2}, \ldots, Y_{in_i})^{\mathrm{T}}$, a covariate ($q \times p$)-dimensional matrix $\mathbf{X}_i^{(m)}$, and an ($n_i \times 1$)-dimensional vector $\mathbf{t}_i=(t_{i1}, t_{i2}, \ldots, t_{in_i})^{\mathrm{T}}$ representing time after starting measurement.
The covariate matrix  $\mathbf{X}_i^{(m)}$ includes genotype data of the $m$-th SNP ($m=1, 2, \ldots, M$).
Because similar analysis is repeated for each of the $M$ SNPs, we shall write $\mathbf{X}_i$ instead of $\mathbf{X}_i^{(m)}$ for simplicity.
Note that the superscript ``$\mathrm{T}$'' indicates the transpose of a matrix or a vector.

In the literature to date, several studies have analyzed a genetic polymorphism in relation to population PK data using NLMMs \cite{Hesselink,Bosch,Bertrand2009,Bertrand2012}.
In NLMMs, it is often assumed that $\mathbf{Y}_i$ arises from the non-linear model,
\[
\mathbf{Y}_i=f(\boldsymbol{\theta}_i, \mathbf{t}_i)+\boldsymbol{\epsilon}_i,
\]
\[
\boldsymbol{\theta}_i=\mathbf{X}_i \boldsymbol{\beta}+\mathbf{Z}_i \boldsymbol{\gamma}_i,
\]
\[
\boldsymbol{\epsilon}_i \sim N(\mathbf{0}, \mathbf{R}_i),
\]
where $f$ is a compartmental model function that is non-linear in its PK parameters $\boldsymbol{\theta}_i$, $\boldsymbol{\epsilon}_i$ is a error vector, $\boldsymbol{\beta}$ and $\boldsymbol{\gamma}_i$ are vectors of fixed effects and random effects, $\mathbf{Z}_i$ is a design matrix for random effects, and $\mathbf{R}_i$  is a covariance matrix.
The $\boldsymbol{\gamma}_i$ are assumed to have a multivariate normal distribution with mean vector $\mathbf{0}$ and a covariance matrix $\mathbf{G}$.
NLMM incorporates unmeasured random effects $\boldsymbol{\gamma}_i$ into the compartmental model function to account for inter-individual variability in the PK parameters $\boldsymbol{\theta}_i$.

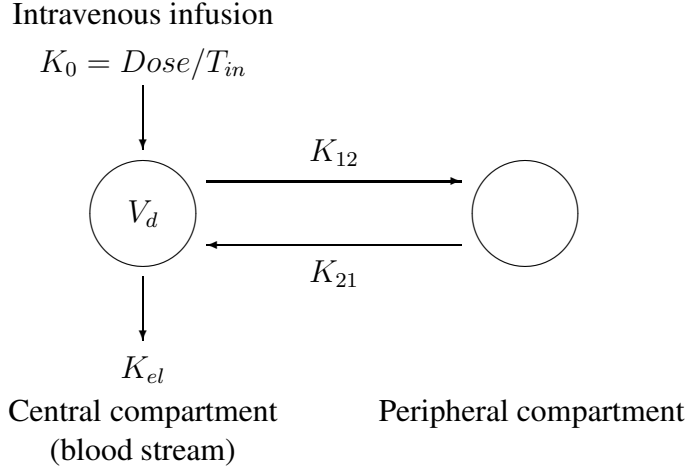
\begin{figure}
\centerline{
\setlength{\unitlength}{.016944mm} 
\begin{picture}(4785,3690)(255,0)
\put(1250,2975){\vector(0,-1){500}}
\put(1250,1475){\vector(0,-1){500}}
\put(1750,2225){\vector(1,0){2000}}
\put(3750,1725){\vector(-1,0){2000}}
\put(1250,1975){\circle{870}}
\put(4250,1975){\circle{870}}
\put(1250,3550){\makebox(0,0)[c]{Intravenous infusion}}
\put(1250,3150){\makebox(0,0)[c]{$K_0=Dose/T_{in}$}}
\put(1250,1975){\makebox(0,0)[c]{$V_{d}$}}
\put(1250,750){\makebox(0,0)[c]{$K_{el}$}}
\put(1250,400){\makebox(0,0)[c]{Central compartment}}
\put(1250,100){\makebox(0,0)[c]{(blood stream)}}
\put(2750,2450){\makebox(0,0)[c]{$K_{12}$}}
\put(2750,1500){\makebox(0,0)[c]{$K_{21}$}}
\put(4300,400){\makebox(0,0)[c]{Peripheral compartment}}
\end{picture}
}
\vspace{12pt}
\caption{
A fitted two-compartment constant intravenous infusion model.
$K_0$ is the infusion rate constant,
$Dose$ is the amount of drug administered,
$T_{in}$ is the infusion time,
$K_{12}$ and $K_{21}$ are inter-compartmental transfer rate constants connecting the central and peripheral compartments,
$V_d$ is the volume of the central compartment,
and $K_{el}$ is the first order elimination rate constant.
}
\end{figure}

Because two-compartment models have been widely used in gemcitabine PK analyses \cite{Scheulen,DePas}, we fitted a two-compartment constant intravenous-infusion model (Figure 1).
For the gemcitabine PGx data, the function form of $f$ is as follows:
\begin{equation}
f(\boldsymbol{\theta}_i, t_{ij})=
\begin{cases}
\begin{matrix}
\frac{K_0^{(i)} (K_{21}^{(i)}-a^{(i)})}{V_d^{(i)} a^{(i)} (a^{(i)}-b^{(i)})}\{\exp(-a^{(i)}t_{ij}) - 1\} + \\ \quad \quad
\frac{K_0^{(i)} (b^{(i)}-K_{21}^{(i)})}{V_d^{(i)} b^{(i)} (a^{(i)}-b^{(i)})}\{\exp(-b^{(i)}t_{ij}) - 1\}
\end{matrix}
& t_{ij} \leq T_{in}^{(i)} \\
\begin{matrix}
\frac{K_0^{(i)} (K_{21}^{(i)}-a^{(i)}) \{\exp(-a^{(i)}T_{in}^{(i)}) - 1\}}{V_d^{(i)} a^{(i)} (a^{(i)}-b^{(i)})}\exp\{-a^{(i)}(t_{ij}-T_{in}^{(i)})\} + \\ \quad \quad
\frac{K_0^{(i)} (b^{(i)}-K_{21}^{(i)}) \{\exp(-b^{(i)}T_{in}^{(i)}) - 1\}}{V_d^{(i)} b^{(i)} (a^{(i)}-b^{(i)})}\exp\{-b^{(i)}(t_{ij}-T_{in}^{(i)})\}
\end{matrix}
& t_{ij} > T_{in}^{(i)}
\end{cases}
\end{equation}
where 
\[
K_0^{(i)}=Dose^{(i)}/T_{in}^{(i)},
\]
\[
a^{(i)}=\{(K_{el}^{(i)}+K_{12}^{(i)}+K_{21}^{(i)} )+\sqrt{(K_{el}^{(i)}+K_{12}^{(i)}+K_{21}^{(i)})^2-4K_{el}^{(i)}K_{21}^{(i)})}\}/2,
\]
\[
b^{(i)}=\{(K_{el}^{(i)}+K_{12}^{(i)}+K_{21}^{(i)} )-\sqrt{(K_{el}^{(i)}+K_{12}^{(i)}+K_{21}^{(i)})^2-4K_{el}^{(i)}K_{21}^{(i)})}\}/2,
\]
and
\[
\boldsymbol{\theta}_i=
\left(
\log V_d^{(i)}, \log K_{el}^{(i)}, \log K_{12}^{(i)}, \log K_{21}^{(i)}
\right)^{\mathrm{T}},
\]
because the PK parameters are restricted to be positive, and in practice are empirically log-transformed \cite{Gabrielsson}.
Commonly, the three genotypes are considered in evaluating the relationship between SNPs and the PK parameters.
We use the dummy variables $x_{iAa}$ and $x_{iAA}$ for the covariate matrix $\mathbf{X}_i$.
Let $(x_{iAa}, x_{iAA})=(0, 0)$, $(1, 0)$, or $(0, 1)$ denote that the $i$-th subject has the genotype ``aa'', ``Aa'', or ``AA'', respectively.
We assume that the effect of a SNP on the PK parameter can be described by the following relationship;
\[
\boldsymbol{\theta}_i=
\begin{pmatrix}
\beta_{V_d}+\beta_{V_{d Aa}}x_{i Aa}+\beta_{V_{d AA}}x_{i AA}+\gamma_{i V_d} \\
\beta_{K_{el}}+\beta_{K_{el Aa}}x_{i Aa}+\beta_{K_{el AA}}x_{i AA}+\gamma_{i K_{el}} \\
\beta_{K_{12}}+\beta_{K_{12 Aa}}x_{i Aa}+\beta_{K_{12 AA}}x_{i AA}+\gamma_{i K_{12}} \\
\beta_{K_{21}}+\beta_{K_{21 Aa}}x_{i Aa}+\beta_{K_{21 AA}}x_{i AA}+\gamma_{i K_{21}}
\end{pmatrix},
\]
where $(\beta_{V_d}, \beta_{K_{el}}, \beta_{K_{12}}, \beta_{K_{21}})$ is an intercept parameter for each PK parameter, and $(\beta_{V_{d Aa}}$, $\beta_{K_{el Aa}}$, $\beta_{K_{12 Aa}}$, $\beta_{K_{21 Aa}})$ and $(\beta_{V_{d AA}}, \beta_{K_{el AA}}, \beta_{K_{12 AA}}, \beta_{K_{21 AA}})$ are the effect parameters of a SNP genotype ``Aa'' and ``AA'' for each PK parameter, respectively.

In order to evaluate the effect parameters of a SNP genotype, a test for the effect for the single genotype (e.g., $H_0: \beta_{V_{d Aa}} = 0$ vs. $H_1: \beta_{V_{d Aa}} \not= 0$) and a multiple degrees-of-freedom test (e.g., $H_0: \beta_{V_{d Aa}} = \beta_{V_{d AA}} = 0$ vs. $H_1: \mbox{not } H_0$) can be considered.
When the null hypotheses, $H_0$, of these tests are rejected at an appropriate significance level, it can be concluded that the SNP affects the profiles of drug concentrations in blood.

However, applying an NLMM to large-scale data, such as genome-wide PGx studies, can be problematic.
(i) NLMMs can be computationally intensive, because these models must compute the marginal log-likelihood by integrating out random effects \cite{Davidian1993}.
(ii) These computations are based on iterative calculations, and may not converge in complex models \cite{Hartford,Lesaffre}.
(iii) Random-effects misspecification leads to bias in parameter estimates of the regression coefficients and slightly inflates type-I error rates of tests for the regression coefficients in an NLMM \cite{Hartford}.

Therefore, we consider an alternative approach that relates to a marginal modeling approach that avoids the specification of random effects.
The approach potentially results in misspecification of the model for the parameters of interest.
Therefore, we evaluated the estimator based on the proposed approach via a misspecified model.


\section{Estimation and inference}

\subsection{Misspecified models}

Misspecified models are useful and powerful tools for studying a behavior of estimators under model misspecification.
Model misspecification means that an incorrect working model is used for estimation.
The proposed method ``intentionally'' uses an incorrect working model that gives an estimator for a weighted average of fixed effect parameter vectors.

The general theory of misspecified models has been proposed by White \cite{White} for maximum likelihood methods, and by Yi and Reid \cite{Yi} for estimating equations.
White showed that the maximum likelihood estimator $\widehat{\boldsymbol{\beta}}$ of a misspecified model converges to a constant vector $\boldsymbol{\beta}_*$ which minimizes the Kullback--Leibler divergence.
A similar property holds in the case of estimating equations.
Under mild conditions, Yi and Reid showed that $\sqrt{K}(\widehat{\boldsymbol{\beta}}-\boldsymbol{\beta}_*)$ is asymptotically normally distributed with a mean vector $\mathbf{0}$ and a covariance matrix $\mathbf{V}_s$ that can be consistently estimated by the so-called sandwich estimator.
However, because $\boldsymbol{\beta}_*$ generally do not have a simple analytical form, we need to evaluate the properties of $\boldsymbol{\beta}_*$.

The literature includes several attempts to uncover the relation between the parameters of a true model and the estimators of model parameters from an incorrect working model.
For example, misspecified models under non-proportional hazards models with a time-varying effect parameter $\beta(t)$ have been discussed in semi-parametric survival models \cite{Xu,Schempter}.
Xu and O'Quigley \cite{Xu} evaluate an asymptotic property of the estimator from a misspecified proportional-hazards model that replaces $\beta(t)$ with a constant $\beta$.
They showed that the estimator $\widehat{\beta}$ converges in probability to a constant $\beta_*$ that is approximated by a weighted average of $\beta(t)$ over time, $\beta_* \approx \int_0^{\infty} \beta(t) v(t) \mathrm{d}F(t) / \int_0^{\infty} v(t) \mathrm{d}F(t)$, where $v(t)$ is the conditional variance of a stochastic process $Z(t)$, and $Z(t)$ is a possibly time-dependent covariate.
Xu and O'Quigley showed that the estimator $\widehat{\beta}$ can be interpreted as a weighted average of true parameters even when an incorrect working model is used.

In this paper, we consider GEE of a misspecified model.
We assume a true model with inter-individual variability in model parameters as fixed-effect parameter vectors $\boldsymbol{\beta}_i$, and a working model that misspecifies the presence of inter-individual variability in model parameters as a common parameter vector $\boldsymbol{\beta}$.
We demonstrate a new interpretation of the estimator $\widehat{\boldsymbol{\beta}}$ as a weighted average of the individual parameter vectors $\boldsymbol{\beta}_i$.

\subsection{Assumptions about the true model}

To describe the true structure of the observations, we assume that the true cumulative distribution of $\mathbf{Y}_i$ is $G(\mathbf{Y}_i; \boldsymbol{\beta}_i, \phi, \boldsymbol{\xi})$ with density $g(\mathbf{Y}_i; \boldsymbol{\beta}_i, \phi, \boldsymbol{\xi})$, where $\boldsymbol{\beta}_i = (\beta_{i1}, \beta_{i2}, \ldots, \beta_{ip})^{\mathrm{T}}$ is a ($p \times 1$)-dimensional vector of effect parameters with inter-individual variability as fixed effects, $\phi$ is a scale parameter, and $\boldsymbol{\xi}$ is a variance model parameter vector.

The expectation of the observation is modeled as $\mathrm{E}[Y_{ij}] = \int Y_{ij} \mathrm{d}G(\mathbf{Y}_i; \boldsymbol{\beta}_i, \phi, \boldsymbol{\xi}) = f(\mathbf{X}_i, t_{ij}; \boldsymbol{\beta}_i)$, where $f$ is a PK function that is nonlinear in its PK parameters $\boldsymbol{\theta}_i$, $\boldsymbol{\theta}_i=(\theta_{i1}, \theta_{i2}, \ldots, \theta_{iq})^{\mathrm{T}}$ is an individual PK parameter ($q \times 1$)-dimensional vector, and $\boldsymbol{\theta}_i$ is modeled in linear form as $\boldsymbol{\theta}_i = \mathbf{X}_i \boldsymbol{\beta}_i$.

Furthermore, we assume that the variance is modeled as $\mathrm{Var}[Y_{ij}] = \phi v(\mathbf{X}_i, t_{ij}; \boldsymbol{\beta}_i, \boldsymbol{\xi})$, where $v$ is a known variance function that has the variance model parameter vector $\boldsymbol{\xi}$.

\subsection{An estimator of a weighted average effect by GEE}

Under the true distribution in Section 3.2, an average effect may be obtained heuristically by replacing $\boldsymbol{\beta}_i$ with a constant $\boldsymbol{\beta}$ and then fitting to data.
We consider GEE of a working model that has parameters $\boldsymbol{\beta}$.
GEE is well known to be inadequate when the mean structure is misspecified.
However, we will show that the estimator $\widehat{\boldsymbol{\beta}}$ can be interpreted as a weighted average effect under the true model, in Section 3.4.

We define GEE of a potentially misspecified model as
\begin{equation}
\mathbf{U}(\boldsymbol{\beta}) =
\sum_{i=1}^K \mathbf{U}_i(\mathbf{Y}_i; \boldsymbol{\beta}) 
=
\sum_{i=1}^K \mathbf{D}_i^{\mathrm{T}} \mathbf{V}_i^{-1} \mathbf{S}_i=
\mathbf{0},
\end{equation}
where
\[
\mathbf{D}_i=
\frac{\partial \boldsymbol{\mu}_i}{\partial \boldsymbol{\beta}^{\mathrm{T}}}=
\left(
\frac{\partial \boldsymbol{\mu}_i}{\partial \beta_1},
\frac{\partial \boldsymbol{\mu}_i}{\partial \beta_2},
\ldots,
\frac{\partial \boldsymbol{\mu}_i}{\partial \beta_p}
\right),
\]
\[
\boldsymbol{\mu}_i=f(\mathbf{X}_i, \mathbf{t}_{i}; \boldsymbol{\beta}),
\]
\[
\mathbf{V}_i=\phi \mathbf{A}_i^{1/2} \mathbf{R}_i(\boldsymbol{\alpha}) \mathbf{A}_i^{1/2},
\]
\[
\mathbf{A}_i=\mathrm{diag}(v(\mathbf{X}_i, t_{i1}; \boldsymbol{\beta}, \boldsymbol{\xi}), v(\mathbf{X}_i, t_{i2}; \boldsymbol{\beta}, \boldsymbol{\xi}), \ldots, v(\mathbf{X}_i, t_{in_i}; \boldsymbol{\beta}, \boldsymbol{\xi})),
\]
\[
\mathbf{S}_i=\mathbf{Y}_i - \boldsymbol{\mu}_i,
\]
$\mathbf{R}_i(\boldsymbol{\alpha})$ is a working correlation ($n_i \times n_i$)-dimensional matrix that can depend on a parameter vector $\boldsymbol{\alpha}$, and $\boldsymbol{\beta}=(\beta_1, \beta_2, \ldots, \beta_p)^{\mathrm{T}}$ is a parameter ($p \times 1$)-dimensional vector that is common between individuals.
The true distribution has the effect parameter vector $\boldsymbol{\beta}_i$, which has inter-individual variability; however, this working model assumes no inter-individual variability.
In equation (2), the individual PK parameter ($q \times 1$)-dimensional vector $\boldsymbol{\psi}_i=(\psi_{i1}, \psi_{i2}, \ldots, \psi_{iq})^{\mathrm{T}}$ is modeled as $\boldsymbol{\psi}_i=\mathbf{X}_i \boldsymbol{\beta}$.
Note that $\mathrm{diag}()$ indicates a diagonal matrix with diagonal elements in parentheses.

Here, we denote the solution to equation (2) as $\widehat{\boldsymbol{\beta}}$.
Yi and Reid \cite{Yi} showed following theorems.
\begin{description}
	\item \textbf{\textit{Theorem 1.}} 
	\textit{Under the true model (the distribution function of $\mathbf{Y}_i$ is $G$), the estimator $\widehat{\boldsymbol{\beta}}$ converges in probability to a constant vector $\boldsymbol{\beta}_*$ as $K \rightarrow \infty$, where $\boldsymbol{\beta}_*$ is a constant vector that satisfies the equation
		\begin{equation}
		\mathrm{E}[\mathbf{U}_i(\mathbf{Y}_i; \boldsymbol{\beta}_*)]=
		\int \mathbf{U}_i(\mathbf{Y}_i; \boldsymbol{\beta}_*) \mathrm{d}G(\mathbf{Y}_i; \boldsymbol{\beta}_i, \phi, \boldsymbol{\xi})=
		\mathbf{0}.
		\end{equation}
	}
	\item \textbf{\textit{Theorem 2.}} 
	\textit{Under the true model, $\sqrt{K}(\widehat{\boldsymbol{\beta}} - \boldsymbol{\beta}_*)$ is asymptotically normal with a mean vector $\mathbf{0}$ and a covariance matrix $\mathbf{V}_s$ as $K \rightarrow \infty$, where 
		\[
		\mathbf{V}_s=\lim_{K \rightarrow \infty} K \mathbf{I}_0(\boldsymbol{\beta}_*)^{-1} \mathbf{I}_1(\boldsymbol{\beta}_*) \mathbf{I}_0(\boldsymbol{\beta}_*)^{-1}
		\]
		is a sandwich variance,
		\[
		\mathbf{I}_0(\boldsymbol{\beta})=
		\mathrm{E} \left[
		\frac{\partial}{\partial \boldsymbol{\beta}}\{ \mathbf{U}(\boldsymbol{\beta}) \}^{\mathrm{T}}
		\right]=
		\sum_{i=1}^K \mathbf{D}_i^{\mathrm{T}} \mathbf{V}_i^{-1} \mathbf{D}_i,
		\]
		and
		\[
		\mathbf{I}_1(\boldsymbol{\beta})=
		\mathrm{E} \left[
		\{ \mathbf{U}(\boldsymbol{\beta}) \}
		\{ \mathbf{U}(\boldsymbol{\beta}) \}^{\mathrm{T}}
		\right]=
		\sum_{i=1}^K \mathbf{D}_i^{\mathrm{T}} \mathbf{V}_i^{-1} \mathrm{Cov} [\mathbf{Y}_i] \mathbf{V}_i^{-1} \mathbf{D}_i.
		\]
	}
\end{description}

\subsection{Interpretation}

The solution $\boldsymbol{\beta}_*$ to equation (3) can be interpreted as a weighted average of the individual parameter vectors $\boldsymbol{\beta}_i$.
In equation (3) from Theorem 1, $\boldsymbol{\beta}_*$ minimizes the distance between the true model and the misspecified model.
For example, when $\mathbf{U}_i$ is the score function, White \cite{White} showed that $\boldsymbol{\beta}_*$ minimizes the Kullback--Leibler divergence between the true model and the misspecified model.

However, $\boldsymbol{\beta}_*$ do not have a simple analytical form.
To evaluate the properties of $\boldsymbol{\beta}_*$, we consider a first-order Taylor expansion of the expectation of equation (2) around $\boldsymbol{\beta}_*=\boldsymbol{\beta}_i$.
As a result, we get
\begin{equation}
\boldsymbol{\beta}_* \approx
\left\{ \sum_{i=1}^K \mathbf{I}_{0i}(\boldsymbol{\beta}_i) \right\}^{-1}
\left\{ \sum_{i=1}^K \mathbf{I}_{0i}(\boldsymbol{\beta}_i) \boldsymbol{\beta}_i \right\}
\end{equation}
which is a weighted average of the individual parameter vectors $\boldsymbol{\beta}_i$ with weights $\mathbf{I}_{0i}(\boldsymbol{\beta}_i)$, where $\mathbf{I}_{0i}(\boldsymbol{\beta}_i)=\mathrm{E}[\partial \mathbf{U}_i(\mathbf{Y}_i; \boldsymbol{\beta}_*) / \partial \boldsymbol{\beta}_*^{\mathrm{T}} \mid_{\boldsymbol{\beta}_*=\boldsymbol{\beta}_i}]$ is an inverse matrix of the model-based variance of the $i$-th subject.
The derivation of equation (4) is shown in Appendix A.

Therefore, according to Theorem 1 and equation (4), the estimator $\widehat{\boldsymbol{\beta}}$ from the working model can asymptotically estimate a weighted average of the individual parameter vectors $\boldsymbol{\beta}_i$.
For instance, if $\boldsymbol{\beta}_i=\boldsymbol{\beta}+\boldsymbol{\gamma}_i$, then $\boldsymbol{\beta}_* \approx \boldsymbol{\beta} + \left\{ \sum_{i=1}^K \mathbf{I}_{0i}(\boldsymbol{\beta}_i) \right\}^{-1}\left\{ \sum_{i=1}^K \mathbf{I}_{0i}(\boldsymbol{\beta}_i) \boldsymbol{\gamma}_i \right\}$;
moreover, if $\boldsymbol{\beta}_i=\boldsymbol{\beta}+\boldsymbol{\gamma}_i$ and  $\boldsymbol{\gamma}_i=\mathbf{0}$, then $\boldsymbol{\beta}_* = \boldsymbol{\beta}$.
The model corresponding to (2) is a misspecified model.
Nevertheless, the estimator $\widehat{\boldsymbol{\beta}}$ converges to $\boldsymbol{\beta}_*$ and can be interpreted as population-weighted average parameters.
Furthermore, this approach is robust for various structures of inter-individual variability (see Section 4.1), because it does not require a strong assumption of a random-effect distribution.

\subsection{
	\texorpdfstring
	{A Wald-type test and an asymptotic $F$-test}
	{A Wald-type test and an asymptotic F-test}
}

As indicated in Section 1, a small--sample correction is needed for valid statistical inference in genome-wide PGx studies that analyze SNP genotyping data.
Therefore, we consider a Wald-type test based on $t$ distributions and an asymptotic $F$-test instead of the asymptotic Wald chi-square tests.

We considered linear hypotheses of the form $H_0: \mathbf{c}^{\mathrm{T}}\boldsymbol{\beta}=0$ versus $H_1: \mathbf{c}^{\mathrm{T}}\boldsymbol{\beta}\not=0$ , where $\mathbf{c}=(c_1, c_2, \ldots, c_p)^{\mathrm{T}}$ is a contrast-coefficient ($p \times 1$)-dimensional vector.

We proposed a Wald-type test statistic;
\[
W=
\frac{\sqrt{K} \mathbf{c}^{\mathrm{T}}\widehat{\boldsymbol{\beta}}}{\sqrt{\mathbf{c}^{\mathrm{T}} \widehat{\mathbf{V}}_s \mathbf{c}}}=
\frac{\sqrt{K} \mathbf{c}^{\mathrm{T}}\widehat{\boldsymbol{\beta}} / \sqrt{\mathbf{c}^{\mathrm{T}} \mathbf{V}_s \mathbf{c}}}{\sqrt{d(\mathbf{c}^{\mathrm{T}} \widehat{\mathbf{V}}_s \mathbf{c} / \mathbf{c}^{\mathrm{T}} \mathbf{V}_s \mathbf{c}) / d}},
\]
where
\begin{equation}
\widehat{\mathbf{V}}_s =
K \widehat{\mathbf{I}}_0(\widehat{\boldsymbol{\beta}})^{-1}
\widehat{\mathbf{I}}_1(\widehat{\boldsymbol{\beta}})
\widehat{\mathbf{I}}_0(\widehat{\boldsymbol{\beta}})^{-1}
\end{equation}
is the estimator of the covariance matrix $\mathbf{V}_s$,
\[
\widehat{\mathbf{I}}_0(\widehat{\boldsymbol{\beta}})=
\sum_{i=1}^K \widehat{\mathbf{D}}_i^{\mathrm{T}} \widehat{\mathbf{V}}_i^{-1} \widehat{\mathbf{D}}_i,
\]
\[
\widehat{\mathbf{I}}_1(\widehat{\boldsymbol{\beta}})=
\sum_{i=1}^K \widehat{\mathbf{D}}_i^{\mathrm{T}} \widehat{\mathbf{V}}_i^{-1} \widehat{\mathrm{Cov}} [\mathbf{Y}_i] \widehat{\mathbf{V}}_i^{-1} \widehat{\mathbf{D}}_i=
\sum_{i=1}^K \widehat{\mathbf{D}}_i^{\mathrm{T}} \widehat{\mathbf{V}}_i^{-1} \widehat{\mathbf{S}}_i \widehat{\mathbf{S}}_i^{\mathrm{T}} \widehat{\mathbf{V}}_i^{-1} \widehat{\mathbf{D}}_i,
\]
\[
\widehat{\mathbf{D}}_i=
\frac{\partial \boldsymbol{\mu}_i}{\partial \boldsymbol{\beta}^{\mathrm{T}}}\bigg|_{\boldsymbol{\beta}=\widehat{\boldsymbol{\beta}}}
\]
\[
\widehat{\boldsymbol{\mu}}_i=f(\mathbf{X}_i, \mathbf{t}_{i}; \widehat{\boldsymbol{\beta}}),
\]
\[
\widehat{\mathbf{V}}_i=\widehat{\phi} \widehat{\mathbf{A}}_i^{1/2} \widehat{\mathbf{R}}_i(\widehat{\boldsymbol{\alpha}}) \widehat{\mathbf{A}}_i^{1/2},
\]
\[
\widehat{\mathbf{A}}_i=\mathrm{diag}(v(\mathbf{X}_i, t_{i1}; \widehat{\boldsymbol{\beta}}, \widehat{\boldsymbol{\xi}}), v(\mathbf{X}_i, t_{i2}; \widehat{\boldsymbol{\beta}}, \widehat{\boldsymbol{\xi}}), \ldots, v(\mathbf{X}_i, t_{in_i}; \widehat{\boldsymbol{\beta}}, \widehat{\boldsymbol{\xi}})),
\]
and
\[
\widehat{\mathbf{S}}_i=\mathbf{y}_i - \widehat{\boldsymbol{\mu}}_i.
\]

Because $\sqrt{K} \mathbf{c}^{\mathrm{T}}\widehat{\boldsymbol{\beta}} / \sqrt{\mathbf{c}^{\mathrm{T}} \mathbf{V}_s \mathbf{c}}$ is asymptotically normally distributed with mean 0 and variance 1 under the null hypothesis, and assuming that $d(\mathbf{c}^{\mathrm{T}} \widehat{\mathbf{V}}_s \mathbf{c} / \mathbf{c}^{\mathrm{T}} \mathbf{V}_s \mathbf{c})$ follows a chi-square distribution with $d$ degrees of freedom (d.f.), the test statistic $W$ is asymptotically $t$-distributed with $d$ d.f., which must be estimated.

We applied the moment estimator of the d.f. $\widehat{d}=\{\mathrm{trace}(\widehat{\boldsymbol{\Psi}}\mathbf{M}) \}^2 / \mathrm{trace}(\widehat{\boldsymbol{\Psi}}\mathbf{M}\widehat{\boldsymbol{\Psi}}\mathbf{M})$, as proposed by Fay and Graubard \cite{Fay}, where $\widehat{\boldsymbol{\Psi}}=\mathrm{block\mbox{-}diag}(\widehat{\boldsymbol{\Psi}}_1, \widehat{\boldsymbol{\Psi}}_2,$ $\ldots,\widehat{\boldsymbol{\Psi}}_K)$ is a block-diagonal matrix, $\widehat{\boldsymbol{\Psi}}_i=\widehat{\mathbf{D}}_i^{\mathrm{T}} \widehat{\mathbf{V}}_i^{-1} \widehat{\mathbf{S}}_i \widehat{\mathbf{S}}_i^{\mathrm{T}} \widehat{\mathbf{V}}_i^{-1} \widehat{\mathbf{D}}_i$, $\mathbf{M}_i=\widehat{\mathbf{D}}_i^{\mathrm{T}} \widehat{\mathbf{V}}_i^{-1} \widehat{\mathbf{D}}_i \mathbf{c} \mathbf{c}^{\mathrm{T}} \widehat{\mathbf{D}}_i \widehat{\mathbf{V}}_i^{-1} \widehat{\mathbf{D}}_i^{\mathrm{T}}$, and $\mathbf{M}=\mathrm{block\mbox{-}diag}(\mathbf{M}_1, \mathbf{M}_2, \ldots, \mathbf{M}_K)$ is a block-diagonal matrix.
We discuss a derivation of $\widehat{d}$ in Appendix B.

We also considered approximating the distribution of multiple degrees-of-freedom tests of $H_0: \mathbf{C}^{\mathrm{T}}\boldsymbol{\beta}=\mathbf{0}$ versus $H_1: \mathbf{C}^{\mathrm{T}}\boldsymbol{\beta}\not=\mathbf{0}$, where $\mathbf{C}=(\mathbf{c}_1, \mathbf{c}_2, \ldots, \mathbf{c}_L)$ is a contrast-coefficient ($p \times L$)-dimensional matrix and $L$ is the number of contrast-coefficient vectors that one wishes to test.

We proposed an asymptotic $F$-test statistic:
\[
F=\frac{1}{L}\left\{
(\mathbf{C}^{\mathrm{T}}\widehat{\boldsymbol{\beta}})^{\mathrm{T}}
(\mathbf{C}^{\mathrm{T}}\widehat{\mathbf{V}}_s\mathbf{C})^{-1}
(\mathbf{C}^{\mathrm{T}}\widehat{\boldsymbol{\beta}})
\right\}.
\]
The statistic $F$ is asymptotically $F$-distributed with a numerator d.f. $L$ and a denominator d.f. $\nu$ that must be estimated.
We applied the moment estimator of the denominator d.f.,
\[
\widehat{\nu}=
\frac{2 \left( \sum_{l=1}^L \frac{\widehat{d}_l}{\widehat{d}_l-2} \right)}{\left( \sum_{l=1}^L \frac{\widehat{d}_l}{\widehat{d}_l-2} \right)-L},
\]
as proposed by Fai and Cornelius \cite{Fai} (see also \cite{Schaalje}), where $\widehat{d}_l$ is the estimator of the d.f. of the Wald-type test statistic for the $l$-th contrast-coefficient vector.

\subsection{
	\texorpdfstring
	{Bias correction for $\widehat{\mathbf{V}}_s$}
	{Bias correction for V_s}
}

The sandwich-variance estimator $\widehat{\mathbf{V}}_s$ is biased downward under small--sample size conditions, as shown by Mancl and DeRouen \cite{Mancl} (see also \cite{MacKinnon,Chesher,Kauermann}).
As indicated in Section 1, a bias correction is also needed in genome-wide PGx studies.

To calculate $\widehat{\mathbf{V}}_s$, a product of the residual vector $\widehat{\mathbf{S}}_i \widehat{\mathbf{S}}_i^{\mathrm{T}}$ is used to estimate $\mathrm{Cov} [\mathbf{Y}_i]$.
However, using a first-order Taylor expansion of $\mathbf{U}(\widehat{\boldsymbol{\beta}})=\mathbf{0}$ and $\widehat{\mathbf{S}}_i$ around $\widehat{\boldsymbol{\beta}}=\boldsymbol{\beta}_*$, $\widehat{\boldsymbol{\beta}}-\boldsymbol{\beta}_* \approx \mathbf{I}_0(\boldsymbol{\beta}_*)^{-1} \sum_{i=1}^K \mathbf{D}_{i*}^{\mathrm{T}} \mathbf{V}_{i*}^{-1} \mathbf{S}_{i*}$ and $\widehat{\mathbf{S}}_i \approx \mathbf{S}_{i*} - \mathbf{D}_{i*}^{\mathrm{T}} (\widehat{\boldsymbol{\beta}}-\boldsymbol{\beta}_*)$, we have $\mathrm{E}[ \widehat{\mathbf{S}}_i \widehat{\mathbf{S}}_i^{\mathrm{T}}] \approx (\mathbf{I}_{n_i} - \mathbf{H}_i) \mathrm{Cov} [\mathbf{Y}_i] (\mathbf{I}_{n_i} - \mathbf{H}_i)^{\mathrm{T}} \not= \mathrm{Cov} [\mathbf{Y}_i]$, where $\mathbf{D}_{i*}$, $\mathbf{V}_{i*}$, and $\mathbf{S}_{i*}$ can be obtained by replacing $\boldsymbol{\beta}$ by $\boldsymbol{\beta}_*$ in the expression $\mathbf{D}_{i}$, $\mathbf{V}_{i}$, and $\mathbf{S}_{i}$; $\mathbf{H}_i=\mathbf{D}_{i*} \mathbf{I}_0(\boldsymbol{\beta}_*)^{-1} \mathbf{D}_{i*}^{\mathrm{T}} \mathbf{V}_{i*}^{-1}$; and $\mathbf{I}_{n_i}$ is an ($n_i \times n_i$)-dimensional identity matrix.
Replacing $\widehat{\mathbf{S}}_i$ in equation (5) by $\widetilde{\mathbf{S}}_i=(\mathbf{I}_{n_i} - \mathbf{H}_i)^{-1}\widehat{\mathbf{S}}_i$ gives the bias-corrected sandwich-variance estimator,
\begin{equation}
\widetilde{\mathbf{V}}_s =
K \widehat{\mathbf{I}}_0(\widehat{\boldsymbol{\beta}})^{-1}
\widetilde{\mathbf{I}}_1(\widehat{\boldsymbol{\beta}})
\widehat{\mathbf{I}}_0(\widehat{\boldsymbol{\beta}})^{-1},
\end{equation}
where
\[
\widetilde{\mathbf{I}}_1(\widehat{\boldsymbol{\beta}})=
\sum_{i=1}^K \widehat{\mathbf{D}}_i^{\mathrm{T}} \widehat{\mathbf{V}}_i^{-1} \widetilde{\mathbf{S}}_i \widetilde{\mathbf{S}}_i^{\mathrm{T}} \widehat{\mathbf{V}}_i^{-1} \widehat{\mathbf{D}}_i,
\]
and $\widehat{\mathbf{H}}_i=\widehat{\mathbf{D}}_i \widehat{\mathbf{I}}_0(\widehat{\boldsymbol{\beta}})^{-1} \widehat{\mathbf{D}}_i^{\mathrm{T}} \widehat{\mathbf{V}}_i^{-1}$ is the leverage of the $i$-th subject \cite{Mancl,Preisser}.
Moreover, an estimator of the d.f. $\widetilde{d}$ and $\widetilde{\nu}$ are given in a similar way;
\[
\widetilde{d}=\{\mathrm{trace}(\widetilde{\boldsymbol{\Psi}}\widetilde{\mathbf{M}}) \}^2 / \mathrm{trace}(\widetilde{\boldsymbol{\Psi}}\widetilde{\mathbf{M}}\widetilde{\boldsymbol{\Psi}}\widetilde{\mathbf{M}})
\]
and
\[
\widetilde{\nu}=\left\{ 2 \left(\sum_{l=1}^L \frac{\widetilde{d}_l}{\widetilde{d}_l-2} \right) \right\} / \left\{ \left(\sum_{l=1}^L \frac{\widetilde{d}_l}{\widetilde{d}_l-2} \right) - L \right\},
\]
where
\[
\widetilde{\boldsymbol{\Psi}}=\mathrm{block\mbox{-}diag}(\widetilde{\boldsymbol{\Psi}}_1, \widetilde{\boldsymbol{\Psi}}_2, \ldots, \widetilde{\boldsymbol{\Psi}}_K),
\]
\[
\widetilde{\boldsymbol{\Psi}}_i=\widehat{\mathbf{D}}_i^{\mathrm{T}} \widehat{\mathbf{V}}_i^{-1} \widetilde{\mathbf{S}}_i \widetilde{\mathbf{S}}_i^{\mathrm{T}} \widehat{\mathbf{V}}_i^{-1} \widehat{\mathbf{D}}_i,\]
\[
\widetilde{\mathbf{M}}=\mathrm{block\mbox{-}diag}(\widetilde{\mathbf{M}}_1, \widetilde{\mathbf{M}}_2, \ldots, \widetilde{\mathbf{M}}_K),
\]
and
\[
\widetilde{\mathbf{M}}_i=(\mathbf{I}_{n_i} - \widehat{\mathbf{H}}_i)^{-1} \widehat{\mathbf{D}}_i^{\mathrm{T}} \widehat{\mathbf{V}}_i^{-1} \widehat{\mathbf{D}}_i \mathbf{c} \mathbf{c}^{\mathrm{T}} \widehat{\mathbf{D}}_i \widehat{\mathbf{V}}_i^{-1} \widehat{\mathbf{D}}_i^{\mathrm{T}} (\mathbf{I}_{n_i} - \widehat{\mathbf{H}}_i)^{-\mathrm{T}}.
\]

\subsection{Example application: Compartment models and the effects of SNPs}

In this study, we applied our methods to a genome-wide PGx study \cite{Sato}, introduced above in Section 2.

We introduce the true distribution of $\mathbf{Y}_i$ and a generalized estimating equation for the data.
It is common to use a constant coefficient of variation (CV) model in PK data analysis \cite{Davidian1995,Wakefield,Sheiner,Beal}.
Under the constant CV model, the expectation and variance of the log-transformed random vector, $\mathbf{Y}_i^*=\log \mathbf{Y}_i$, is modeled as $\mathrm{E}[Y_{ij}^*]=f^*(\mathbf{X}_i, t_{ij}; \boldsymbol{\beta}_i)$, $\mathrm{Var}[Y_{ij}^*]=\sigma^2$, and $v(\mathbf{X}_i, t_{ij}; \boldsymbol{\beta}_i, \boldsymbol{\xi})=1$, where
\begin{equation}
f^*(\mathbf{X}_i, \mathbf{t}_{i}; \boldsymbol{\beta}_i)=
\log f(\mathbf{X}_i, \mathbf{t}_{i}; \boldsymbol{\beta}_i)
\end{equation}
is the log-transformed PK function, and $f(\mathbf{X}_i, \mathbf{t}_{i}; \boldsymbol{\beta}_i)$ is given by equation (1).
Furthermore, the working correlation matrix is modeled as, $\mathbf{R}_i(\boldsymbol{\alpha})=\mathbf{I}_{n_i}$.

Along with NLMM as shown in Section 2, in order to evaluate the association between PK parameters and SNPs, the individual PK parameters in GEE of a working model is modeled as 
\[
\boldsymbol{\psi}_i=
\begin{pmatrix}
\log V_d^{(i)} \\
\log K_{el}^{(i)} \\
\log K_{12}^{(i)} \\
\log K_{21}^{(i)} \\
\end{pmatrix}
=
\mathbf{X}_i \boldsymbol{\beta}
=
\begin{pmatrix}
\beta_{V_d} + \beta_{V_d Aa} x_{i Aa} + \beta_{V_d AA} x_{i AA} \\
\beta_{K_{el}} + \beta_{K_{el} Aa} x_{i Aa} + \beta_{K_{el} AA} x_{i AA} \\
\beta_{K_{12}} + \beta_{K_{12} Aa} x_{i Aa} + \beta_{K_{12} AA} x_{i AA} \\
\beta_{K_{21}} + \beta_{K_{21} Aa} x_{i Aa} + \beta_{K_{21} AA} x_{i AA} \\
\end{pmatrix},
\]
where the covariate matrix is
\makeatletter
\c@MaxMatrixCols=12
\makeatother
\begin{equation}
\mathbf{X}_i = 
\begin{pmatrix}
1 & x_{i Aa} & x_{i AA} & 0 & 0 & 0 & 0 & 0 & 0 & 0 & 0 & 0 \\
0 & 0 & 0 & 1 & x_{i Aa} & x_{i AA} & 0 & 0 & 0 & 0 & 0 & 0 \\
0 & 0 & 0 & 0 & 0 & 0 & 1 & x_{i Aa} & x_{i AA} & 0 & 0 & 0 \\
0 & 0 & 0 & 0 & 0 & 0 & 0 & 0 & 0 & 1 & x_{i Aa} & x_{i AA} \\
\end{pmatrix},
\end{equation}
and the parameter vector is 
\[
\boldsymbol{\beta}=(
\beta_{V_d}, \beta_{V_d Aa}, \beta_{V_d AA},
\beta_{K_{el}}, \beta_{K_{el} Aa}, \beta_{K_{el} AA},
\beta_{K_{12}}, \beta_{K_{12} Aa}, \beta_{K_{12} AA},
\beta_{K_{21}}, \beta_{K_{21} Aa}, \beta_{K_{21} AA}
)^{\mathrm{T}}.
\]

\section{Simulations}

Simulations were conducted to study the performance of the proposed Wald-type test and asymptotic $F$-test for population PK data.
The simulation conditions for population PK data were determined by reference to an actual genome-wide PGx study \cite{Sato}.

For simplicity, observed responses $\mathbf{y}_i^*$, which should have individual variations, were generated from the NLMM as follows:
\begin{equation}
\boldsymbol{\theta}_i=
\begin{pmatrix}
(\beta_{V_d} + \gamma_{i V_d}) + \beta_{V_d Aa} x_{i Aa} + \beta_{V_d AA} x_{i AA} \\
\beta_{K_{el}} + \beta_{K_{el} Aa} x_{i Aa} + \beta_{K_{el} AA} x_{i AA} \\
(\beta_{K_{12}} + \gamma_{i K_{12}}) + \beta_{K_{12} Aa} x_{i Aa} + \beta_{K_{12} AA} x_{i AA} \\
(\beta_{K_{21}} + \gamma_{i K_{21}}) + \beta_{K_{21} Aa} x_{i Aa} + \beta_{K_{21} AA} x_{i AA} \\
\end{pmatrix}
\end{equation}
where $\boldsymbol{\gamma}_i=(\gamma_{i V_d}, \gamma_{i K_{12}}, \gamma_{i K_{21}})^{\mathrm{T}}$ is a random-effect vector of the $i$-th subject for each PK parameter, for which conditions are shown in Section 4.1 and Section 4.2.
Further, we assumed $\mathbf{Y}_i^* | \boldsymbol{\gamma}_i \sim \mathrm{N}(f^*(\mathbf{X}_i, \mathbf{t}_{i}; \boldsymbol{\beta}_i), \sigma^2 \mathbf{I}_{n_i})$, where $f^*$ is a log-transformed two-compartment constant intravenous-infusion PK function in equation (7) setting $Dose^{(i)}$ to 1400 mg and $T_{in}^{(i)}$ to 0.5 hours.
The intercept terms of the log-transformed PK parameters were set to $\beta_{V_d}=3.72$, $\beta_{K_{el}}=1.38$, $\beta_{K_{12}}=-1.89$, and $\beta_{K_{21}}=-0.35$;
the standard deviation $\sigma$ was set to 0.27;
and values of the remaining parameters are shown in Section 4.1 and Section 4.2.
Note that these parameters were set based on a preliminary NLMM analysis of gemcitabine data without covariates;
we assumed that the random-effect vector $\boldsymbol{\gamma}_i$ is normally distributed with a mean vector $\mathbf{0}$ and a diagonal covariance matrix $\mathrm{diag}(\tau_{V_d}^2, \tau_{K_{12}}^2, \tau_{K_{21}}^2)$, because the random effects of the elimination parameter $K_{el}$ were too small.
In the simulations, we changed the blood sampling points to 0.1, 0.5, 0.75, 1.0, 1.5, 2.0, 2.5, and 4.5 hours after drug administration.
The covariate matrices $\mathbf{X}_i$ in equation (8) were generated non-randomly by taking
\[
(x_{i Aa}, x_{i AA})=
\begin{cases}
(0, 0) & \mbox{genotype aa} \\
(1, 0) & \mbox{genotype Aa} \\
(0, 1) & \mbox{genotype AA} \\
\end{cases}.
\]

In actual studies, the sample size of each genotype group is not controlled, but depends on allele frequency.
Generally, these studies are likely to have unequal sample sizes for different genotypes and a minor-allele frequency (MAF) of less than 0.5, most commonly around 0.2 \cite{Hirakawa}.
When MAF is small, there are too few subjects homozygous for the minor allele.
Therefore, the MAF was set to 0.25 or 0.50.
If we let $n_{aa}$, $n_{Aa}$, and $n_{AA}$ denote the sample size of each genotype, then $\mbox{MAF} = (n_{Aa}+2n_{AA})/2n$.
We assumed that the population was in Hardy--Weinberg equilibrium, with the total sample size $n=100$;
the sample size for each group was set to $n_{aa}=56$, $n_{Aa}=37$, and $n_{AA}=7$ for $\mbox{MAF} = 0.25$, and $n_{aa}=25$, $n_{Aa}=50$, and $n_{AA}=25$ for $\mbox{MAF} = 0.50$.
The total sample size $n=100$ is not realistic for genome-wide PGx studies, but is sufficient to evaluate statistical performance.

For each data configuration, 1000 simulations were generated.

For each simulation, the generalized estimating equation of a misspecified model in equation (2) was fitted assuming the two-compartment constant intravenous-infusion model as shown in Section 3.7, and the NLMM was fitted assuming a normal random-effects model with adaptive Gauss--Hermite quadrature.
We used a diagonal covariance matrix for the random-effect vector $\boldsymbol{\gamma}_i$, which is normally distributed with a mean vector $\mathbf{0}$, and a covariance matrix $\mathrm{diag}(\tau_{V_d}^2, \tau_{K_{12}}^2, \tau_{K_{21}}^2)$ for the NLMM, which is a commonly-used method (see Section 2).

In order to assess the statistical performance of tests for the effect of a SNP on PK parameters (e.g., $H_0: \beta_{V_d Aa}  = 0$ vs.  $H_1: \beta_{V_d Aa} \not= 0$ and $H_0: \beta_{V_d Aa}=\beta_{V_d AA}=0$ vs.  $H_1: \mbox{not } H_0$), we applied the proposed Wald-type test using $\widehat{\mathbf{V}}_s$ in equation (5) (hereinafter referred to as GEE ($\widehat{\mathbf{V}}_s$)) and $\widetilde{\mathbf{V}}_s$ in equation (6) (hereinafter referred to as GEE ($\widetilde{\mathbf{V}}_s$)), and a Wald test in the NLMM for testing linear hypotheses.
In addition, we applied the proposed asymptotic $F$-test using $\widehat{\mathbf{V}}_s$ in equation (5) and $\widetilde{\mathbf{V}}_s$ in equation (6), and an asymptotic $F$-test in the NLMM for testing linear hypotheses.
As we tested for the effect of the SNP on the parameters $V_d$, $K_{el}$, $K_{12}$, and $K_{21}$ as a whole, we used $L=2$,
\[
\mathbf{C}_{V_d} = 
\begin{pmatrix}
0 & 1 & 0 & 0 & 0 & 0 & 0 & 0 & 0 & 0 & 0 & 0 \\
0 & 0 & 1 & 0 & 0 & 0 & 0 & 0 & 0 & 0 & 0 & 0 \\
\end{pmatrix},
\]
\[
\mathbf{C}_{K_{el}} = 
\begin{pmatrix}
0 & 0 & 0 & 0 & 1 & 0 & 0 & 0 & 0 & 0 & 0 & 0 \\
0 & 0 & 0 & 0 & 0 & 1 & 0 & 0 & 0 & 0 & 0 & 0 \\
\end{pmatrix},
\]
\[
\mathbf{C}_{K_{12}} = 
\begin{pmatrix}
0 & 0 & 0 & 0 & 0 & 0 & 0 & 1 & 0 & 0 & 0 & 0 \\
0 & 0 & 0 & 0 & 0 & 0 & 0 & 0 & 1 & 0 & 0 & 0 \\
\end{pmatrix},
\]
and
\[
\mathbf{C}_{K_{21}} = 
\begin{pmatrix}
0 & 0 & 0 & 0 & 0 & 0 & 0 & 0 & 0 & 0 & 1 & 0 \\
0 & 0 & 0 & 0 & 0 & 0 & 0 & 0 & 0 & 0 & 0 & 1 \\
\end{pmatrix},
\]
respectively.
Note that the simulation data assume that the effect parameters of a SNP do not include random effects (e.g., $\beta_{V_d Aa*} \approx \beta_{V_d Aa}$) as shown in equation (9).
The two-tailed significance level of all tests was set to 0.05.

For each simulation, we evaluated the type-I error rates, powers, mean biases, mean-squared errors (MSEs), computation times, and convergence proportions of the iterative calculations.
Furthermore, we implemented numerical computations for estimation and inference of GEE ($\widehat{\mathbf{V}}_s$) and GEE ($\widetilde{\mathbf{V}}_s$) using the SAS/NLMIXED Procedure (version 9.2, SAS Institute, Inc., Cary, North Carolina).

\subsection{Type-I error rates}
In this Section, we consider whether GEE ($\widehat{\mathbf{V}}_s$), GEE ($\widetilde{\mathbf{V}}_s$), and NLMM can control type-I error rates under random-effects misspecification in the following scenarios:
\begin{description}
	\item \textbf{Scenario 1} 
	The random-effects distributions are ``correctly specified'' in NLMM.
	We assumed that the random-effect vector $\boldsymbol{\gamma}_i$ is normally distributed with a mean vector $\mathbf{0}$ and a covariance matrix $\mathrm{diag}(\tau_{V_d}^2, \tau_{K_{12}}^2, \tau_{K_{21}}^2)$, where the standard deviation of random effects was set to 0.12, 0.68, and 0.89 based on the preliminary analysis.
	\item \textbf{Scenario 2} 
	The random-effects distributions are ``misspecified'' in NLMM.
	We assumed that each element of the random-effect vector $\boldsymbol{\gamma}_i$ follows a uniform distribution;
	\[
	\gamma_{i V_d} \sim \mathrm{Uniform}(-\frac{1}{2}\sqrt{12\tau_{V_d}^2},\frac{1}{2}\sqrt{12\tau_{V_d}^2}),
	\]
	\[
	\gamma_{i K_{12}} \sim \mathrm{Uniform}(-\frac{1}{2}\sqrt{12\tau_{K_{12}}^2}, \frac{1}{2}\sqrt{12\tau_{K_{12}}^2}),
	\]
	and
	\[
	\gamma_{i K_{21}} \sim \mathrm{Uniform}(-\frac{1}{2}\sqrt{12\tau_{K_{21}}^2}, \frac{1}{2}\sqrt{12\tau_{K_{21}}^2}),
	\]
	where the random-effect parameters $\tau_{V_d}$, $\tau_{K_{12}}$, and $\tau_{K_{21}}$ are set to 0.12, 0.68, and 0.89, respectively.
	Here, $\mathrm{Var}[\gamma_{i \mbox{\textbullet}}]=\tau_{\mbox{\textbullet}}^2$.
	\item \textbf{Scenario 3} 
	The random-effects distributions are ``misspecified'' in NLMM.
	We assumed that each element of the random-effect vector $\boldsymbol{\gamma}_i$ follows a gamma distribution;
	$\gamma_{i V_d} \sim \mathrm{Gamma}(\tau_{V_d}^2, 1)$, $\gamma_{i K_{12}} \sim \mathrm{Gamma}(\tau_{K_{12}}^2, 1)$, and $\gamma_{i K_{21}} \sim \mathrm{Gamma}(\tau_{K_{21}}^2, 1)$, where the random-effect parameters $\tau_{V_d}$, $\tau_{K_{12}}$, and $\tau_{K_{21}}$ are set to 0.12, 0.68, and 0.89, respectively.
	Here, $\mathrm{Var}[\gamma_{i \mbox{\textbullet}}]=\tau_{\mbox{\textbullet}}^2$.
\end{description}

To evaluate type-I error rates, parameters of the SNP effect $\beta_{V_d Aa}$, $\beta_{V_d AA}$,  $\beta_{K_{el} Aa}$, $\beta_{K_{el} AA}$, $\beta_{K_{12} Aa}$, $\beta_{K_{12} AA}$, $\beta_{K_{21} Aa}$, and $\beta_{K_{21} AA}$ are set to 0.0.

Type-I error rates, biases, and MSEs of GEE ($\widehat{\mathbf{V}}_s$), GEE ($\widetilde{\mathbf{V}}_s$), and NLMM for MAF = 0.25 and 0.50 are shown in Tables 1 and 2.
Because GEE ($\widehat{\mathbf{V}}_s$) and GEE ($\widetilde{\mathbf{V}}_s$) differ only in variances, results of biases and MSEs were combined as GEE ($\widehat{\mathbf{V}}_s$, $\widetilde{\mathbf{V}}_s$) in Tables 1 and 2.

\begin{table}
	\small
	\caption{Type-I errors, biases, and MSEs of GEE ($\widehat{\mathbf{V}}_s$), GEE ($\widetilde{\mathbf{V}}_s$), and NLMM for MAF = 0.25 in Scenarios~1--3.}
	\label{t:typeIError}
	\centering
	\begin{tabular}{lcrrrrrrrr}\hline
		\multicolumn{2}{c}{Parameter} & \multicolumn{1}{c}{$\beta_{V_dAa}$} & \multicolumn{1}{c}{$\beta_{V_dAA}$} & \multicolumn{1}{c}{$\beta_{K_{el}Aa}$} & \multicolumn{1}{c}{$\beta_{K_{el}AA}$} & \multicolumn{1}{c}{$\beta_{K_{12}Aa}$} & \multicolumn{1}{c}{$\beta_{K_{12}AA}$} & \multicolumn{1}{c}{$\beta_{K_{21}Aa}$} & \multicolumn{1}{c}{$\beta_{K_{21}AA}$} \\ \hline

		\multicolumn{10}{c}{Type-I error rate} \\ \hline
		\multicolumn{10}{l}{Scenario 1: ``correctly specified'' in NLMM} \\
		GEE ($\widehat{\mathbf{V}}_s$) &  & 0.047 & 0.046 & 0.044 & 0.049 & 0.038 & 0.052 & 0.053 & 0.054 \\
		GEE ($\widetilde{\mathbf{V}}_s$) & \multicolumn{1}{c}{Wald} & 0.047 & 0.043 & 0.043 & 0.041 & 0.037 & 0.048 & 0.050 & 0.050 \\
		NLMM$^{\mathrm{a}}$ &  & 0.058 & 0.051 & 0.061 & 0.040 & 0.050 & 0.056 & \llap{$^\mathrm{c}$}\textbf{0.067} & 0.065 \\
		GEE ($\widehat{\mathbf{V}}_s$) &  & \multicolumn{2}{c}{\llap{$^\mathrm{c}$}\textbf{0.030}} & \multicolumn{2}{c}{0.041} & \multicolumn{2}{c}{0.039} & \multicolumn{2}{c}{\llap{$^\mathrm{c}$}\textbf{0.032}} \\
		GEE ($\widetilde{\mathbf{V}}_s$) & \multicolumn{1}{c}{$F$} & \multicolumn{2}{c}{\llap{$^\mathrm{c}$}\textbf{0.024}} & \multicolumn{2}{c}{\llap{$^\mathrm{c}$}\textbf{0.033}} & \multicolumn{2}{c}{\llap{$^\mathrm{c}$}\textbf{0.028}} & \multicolumn{2}{c}{\llap{$^\mathrm{c}$}\textbf{0.019}} \\
		NLMM$^{\mathrm{a}}$ &  & \multicolumn{2}{c}{\llap{$^\mathrm{c}$}\textbf{0.074}} & \multicolumn{2}{c}{\llap{$^\mathrm{c}$}\textbf{0.094}} & \multicolumn{2}{c}{0.052} & \multicolumn{2}{c}{\llap{$^\mathrm{c}$}\textbf{0.077}} \\
		\multicolumn{10}{l}{Scenario 2: ``misspecified'' in NLMM (Uniform)} \\
		GEE ($\widehat{\mathbf{V}}_s$) &  & 0.040 & 0.040 & 0.041 & 0.045 & 0.053 & 0.043 & 0.040 & 0.045 \\
		GEE ($\widetilde{\mathbf{V}}_s$) & \multicolumn{1}{c}{Wald} & 0.038 & \llap{$^\mathrm{c}$}\textbf{0.034} & 0.037 & \llap{$^\mathrm{c}$}\textbf{0.034} & 0.049 & \llap{$^\mathrm{c}$}\textbf{0.030} & 0.038 & \llap{$^\mathrm{c}$}\textbf{0.027} \\
		NLMM$^{\mathrm{a}}$ &  & 0.058 & 0.064 & \llap{$^\mathrm{c}$}\textbf{0.075} & 0.052 & 0.064 & 0.056 & \llap{$^\mathrm{c}$}\textbf{0.078} & \llap{$^\mathrm{c}$}\textbf{0.068} \\
		GEE ($\widehat{\mathbf{V}}_s$) &  & \multicolumn{2}{c}{0.036} & \multicolumn{2}{c}{0.035} & \multicolumn{2}{c}{0.038} & \multicolumn{2}{c}{0.035} \\
		GEE ($\widetilde{\mathbf{V}}_s$) & \multicolumn{1}{c}{$F$} & \multicolumn{2}{c}{\llap{$^\mathrm{c}$}\textbf{0.032}} & \multicolumn{2}{c}{\llap{$^\mathrm{c}$}\textbf{0.026}} & \multicolumn{2}{c}{\llap{$^\mathrm{c}$}\textbf{0.029}} & \multicolumn{2}{c}{\llap{$^\mathrm{c}$}\textbf{0.021}} \\
		NLMM$^{\mathrm{a}}$ &  & \multicolumn{2}{c}{0.061} & \multicolumn{2}{c}{\llap{$^\mathrm{c}$}\textbf{0.071}} & \multicolumn{2}{c}{0.055} & \multicolumn{2}{c}{\llap{$^\mathrm{c}$}\textbf{0.086}} \\
		\multicolumn{10}{l}{Scenario 3: ``misspecified'' in NLMM (Gamma)} \\
		GEE ($\widehat{\mathbf{V}}_s$) &  & \llap{$^\mathrm{c}$}\textbf{0.025} & 0.042 & \llap{$^\mathrm{c}$}\textbf{0.034} & \llap{$^\mathrm{c}$}\textbf{0.033} & 0.042 & 0.054 & 0.048 & 0.037 \\
		GEE ($\widetilde{\mathbf{V}}_s$) & \multicolumn{1}{c}{Wald} & \llap{$^\mathrm{c}$}\textbf{0.023} & 0.039 & \llap{$^\mathrm{c}$}\textbf{0.032} & \llap{$^\mathrm{c}$}\textbf{0.019} & 0.041 & 0.041 & 0.041 & \llap{$^\mathrm{c}$}\textbf{0.025} \\
		NLMM$^{\mathrm{a}}$ &  & 0.035 & 0.036 & \llap{$^\mathrm{c}$}\textbf{0.068} & 0.059 & \llap{$^\mathrm{c}$}\textbf{0.026} & \llap{$^\mathrm{c}$}\textbf{0.033} & 0.062 & 0.057 \\
		GEE ($\widehat{\mathbf{V}}_s$) &  & \multicolumn{2}{c}{0.037} & \multicolumn{2}{c}{\llap{$^\mathrm{c}$}\textbf{0.027}} & \multicolumn{2}{c}{0.042} & \multicolumn{2}{c}{\llap{$^\mathrm{c}$}\textbf{0.034}} \\
		GEE ($\widetilde{\mathbf{V}}_s$) & \multicolumn{1}{c}{$F$} & \multicolumn{2}{c}{\llap{$^\mathrm{c}$}\textbf{0.031}} & \multicolumn{2}{c}{\llap{$^\mathrm{c}$}\textbf{0.023}} & \multicolumn{2}{c}{0.036} & \multicolumn{2}{c}{\llap{$^\mathrm{c}$}\textbf{0.026}} \\
		NLMM$^{\mathrm{a}}$ &  & \multicolumn{2}{c}{0.043} & \multicolumn{2}{c}{0.062} & \multicolumn{2}{c}{\llap{$^\mathrm{c}$}\textbf{0.029}} & \multicolumn{2}{c}{\llap{$^\mathrm{c}$}\textbf{0.068}} \\ \hline
		
		\multicolumn{10}{c}{Bias} \\ \hline
		\multicolumn{10}{l}{Scenario 1: ``correctly specified'' in NLMM} \\
		\multicolumn{2}{l}{GEE ($\widehat{\mathbf{V}}_s$, $\widetilde{\mathbf{V}}_s$)$^{\mathrm{b}}$} & \llap{$-$}0.002 & \llap{$-$}0.002 & \llap{$<$}0.001 & \llap{$-$}0.001 & 0.012 & 0.008 & \llap{$-$}0.005 & \llap{$-$}0.055 \\
		\multicolumn{2}{l}{NLMM$^{\mathrm{a}}$} & 0.003 & 0.001 & \llap{$<$}0.001 & \llap{$-$}0.004 & 0.011 & \llap{$-$}0.017 & \llap{$-$}0.005 & \llap{$-$}0.050 \\
		\multicolumn{10}{l}{Scenario 2: ``misspecified'' in NLMM (Uniform)} \\
		\multicolumn{2}{l}{GEE ($\widehat{\mathbf{V}}_s$, $\widetilde{\mathbf{V}}_s$)$^{\mathrm{b}}$} & \llap{$-$}0.001 & \llap{$-$}0.006 & \llap{$<$}0.001 & 0.002 & \llap{$-$}0.004 & \llap{$^{\mathrm{d}}$}\textbf{0.033} & \llap{$-$}0.007 & \llap{$^{\mathrm{d}}\mathbf{-}$}\textbf{0.039} \\
		\multicolumn{2}{l}{NLMM$^{\mathrm{a}}$} & \llap{$<$}0.001 & 0.003 & \llap{$-$}0.001 & \llap{$-$}0.001 & \llap{$-$}0.002 & \llap{$-$}0.005 & 0.003 & 0.001 \\
		\multicolumn{10}{l}{Scenario 3: ``misspecified'' in NLMM (Gamma)} \\
		\multicolumn{2}{l}{GEE ($\widehat{\mathbf{V}}_s$, $\widetilde{\mathbf{V}}_s$)$^{\mathrm{b}}$} & 0.001 & \llap{$-$}0.001 & \llap{$-$}0.001 & 0.002 & \llap{$-$}0.011 & \llap{$^{\mathrm{d}}\mathbf{-}$}\textbf{0.019} & \llap{$-$}0.004 & \llap{$-$}0.033 \\
		\multicolumn{2}{l}{NLMM$^{\mathrm{a}}$} & \llap{$-$}0.001 & 0.005 & \llap{$<$}0.001 & \llap{$-$}0.004 & \llap{$-$}0.002 & 0.005 & \llap{$-$}0.002 & \llap{$-$}0.028 \\ \hline
		
		\multicolumn{10}{c}{MSE} \\ \hline
		\multicolumn{10}{l}{Scenario 1: ``correctly specified'' in NLMM} \\
		\multicolumn{2}{l}{GEE ($\widehat{\mathbf{V}}_s$, $\widetilde{\mathbf{V}}_s$)$^{\mathrm{b}}$} & 0.003 & 0.009 & \llap{$<$}0.001 & 0.002 & 0.033 & \llap{$^{\mathrm{d}}$}\textbf{0.124} & 0.041 & 0.160 \\
		\multicolumn{2}{l}{NLMM$^{\mathrm{a}}$} & 0.002 & 0.009 & \llap{$<$}0.001 & 0.002 & 0.027 & 0.097 & 0.047 & 0.166 \\
		\multicolumn{10}{l}{Scenario 2: ``misspecified'' in NLMM (Uniform)} \\
		\multicolumn{2}{l}{GEE ($\widehat{\mathbf{V}}_s$, $\widetilde{\mathbf{V}}_s$)$^{\mathrm{b}}$} & 0.002 & 0.009 & \llap{$<$}0.001 & 0.002 & 0.034 & \llap{$^{\mathrm{d}}$}\textbf{0.121} & 0.033 & 0.132 \\
		\multicolumn{2}{l}{NLMM$^{\mathrm{a}}$} & 0.003 & 0.009 & \llap{$<$}0.001 & 0.001 & 0.028 & 0.104 & \llap{$^{\mathrm{d}}$}\textbf{0.046} & \llap{$^{\mathrm{d}}$}\textbf{0.164} \\
		\multicolumn{10}{l}{Scenario 3: ``misspecified'' in NLMM (Gamma)} \\
		\multicolumn{2}{l}{GEE ($\widehat{\mathbf{V}}_s$, $\widetilde{\mathbf{V}}_s$)$^{\mathrm{b}}$} & 0.002 & 0.008 & \llap{$<$}0.001 & 0.002 & 0.031 & \llap{$^{\mathrm{d}}$}\textbf{0.097} & 0.015 & 0.057 \\
		\multicolumn{2}{l}{NLMM$^{\mathrm{a}}$} & 0.003 & 0.008 & \llap{$<$}0.001 & 0.002 & 0.021 & 0.071 & 0.018 & 0.062 \\ \hline
		
		\multicolumn{10}{p{15.9cm}}{\scriptsize $^{\mathrm{a}}$Values of the NLMM have been calculated from simulations with a low convergence proportion for iterative calculation (see Subsection 3.3).} \\
		\multicolumn{10}{p{15.9cm}}{\scriptsize $^{\mathrm{b}}$Because GEE ($\widehat{\mathbf{V}}_s$) and GEE ($\widetilde{\mathbf{V}}_s$) differ only in variances, results of biases were combined as GEE ($\widehat{\mathbf{V}}_s$, $\widetilde{\mathbf{V}}_s$).} \\
		\multicolumn{10}{p{15.9cm}}{\scriptsize $^{\mathrm{c}}$Values that were $\leq 0.034$ or $\geq 0.066$ (binomial 99 \% upper confidential limit, $0.05\pm2.33\sqrt{0.05(1-0.05)/1000}$) are highlighted.} \\
		\multicolumn{10}{p{15.9cm}}{\scriptsize $^{\mathrm{d}}$Values for which there was more than a 0.01 inferior difference between GEE ($\widehat{\mathbf{V}}_s$, $\widetilde{\mathbf{V}}_s$) and the NLMM are highlighted.}\\
		\multicolumn{10}{p{15.9cm}}{\scriptsize The number of simulations was 1000.} \\

	\end{tabular}
\end{table}

\begin{table}
	\small
	\caption{Type-I errors, biases, and MSEs of GEE ($\widehat{\mathbf{V}}_s$), GEE ($\widetilde{\mathbf{V}}_s$), and NLMM for MAF = 0.50 in Scenarios~1--3.}
	\label{t:typeIError050}
	\centering
	\begin{tabular}{lcrrrrrrrr} \hline
		\multicolumn{2}{c}{Parameter} & \multicolumn{1}{c}{$\beta_{V_dAa}$} & \multicolumn{1}{c}{$\beta_{V_dAA}$} & \multicolumn{1}{c}{$\beta_{K_{el}Aa}$} & \multicolumn{1}{c}{$\beta_{K_{el}AA}$} & \multicolumn{1}{c}{$\beta_{K_{12}Aa}$} & \multicolumn{1}{c}{$\beta_{K_{12}AA}$} & \multicolumn{1}{c}{$\beta_{K_{21}Aa}$} & \multicolumn{1}{c}{$\beta_{K_{21}AA}$} \\ \hline
		
		\multicolumn{10}{c}{Type-I error rate} \\ \hline
		\multicolumn{10}{l}{Scenario 1: ``correctly specified'' in NLMM} \\
		GEE ($\widehat{\mathbf{V}}_s$) &  & 0.036 & 0.040 & 0.052 & 0.040 & 0.049 & 0.054 & 0.044 & 0.043 \\
		GEE ($\widetilde{\mathbf{V}}_s$) & \multicolumn{1}{c}{Wald} & 0.035 & 0.038 & 0.051 & 0.036 & 0.047 & 0.052 & 0.044 & 0.037 \\
		NLMM$^{\mathrm{a}}$ &  & 0.044 & 0.051 & 0.054 & 0.061 & \llap{$^\mathrm{c}$}\textbf{0.067} & 0.061 & \llap{$^\mathrm{c}$}\textbf{0.073} & 0.058 \\
		GEE ($\widehat{\mathbf{V}}_s$) &  & \multicolumn{2}{c}{0.061} & \multicolumn{2}{c}{0.047} & \multicolumn{2}{c}{0.052} & \multicolumn{2}{c}{0.038} \\
		GEE ($\widetilde{\mathbf{V}}_s$) & \multicolumn{1}{c}{$F$} & \multicolumn{2}{c}{0.058} & \multicolumn{2}{c}{0.040} & \multicolumn{2}{c}{0.047} & \multicolumn{2}{c}{\llap{$^\mathrm{c}$}\textbf{0.030}} \\
		NLMM$^{\mathrm{a}}$ &  & \multicolumn{2}{c}{\llap{$^\mathrm{c}$}\textbf{0.072}} & \multicolumn{2}{c}{\llap{$^\mathrm{c}$}\textbf{0.076}} & \multicolumn{2}{c}{\llap{$^\mathrm{c}$}\textbf{0.080}} & \multicolumn{2}{c}{\llap{$^\mathrm{c}$}\textbf{0.089}} \\
		\multicolumn{10}{l}{Scenario 2: ``misspecified'' in NLMM (Uniform)} \\
		GEE ($\widehat{\mathbf{V}}_s$) &  & 0.037 & 0.038 & 0.046 & 0.038 & 0.037 & 0.041 & 0.030 & 0.034 \\
		GEE ($\widetilde{\mathbf{V}}_s$) & \multicolumn{1}{c}{Wald} & 0.037 & 0.035 & 0.043 & 0.035 & 0.033 & 0.036 & 0.028 & 0.031 \\
		NLMM$^{\mathrm{a}}$ &  & 0.053 & 0.047 & \llap{$^\mathrm{c}$}\textbf{0.080} & \llap{$^\mathrm{c}$}\textbf{0.068} & 0.062 & 0.060 & \llap{$^\mathrm{c}$}\textbf{0.077} & \llap{$^\mathrm{c}$}\textbf{0.068} \\
		GEE ($\widehat{\mathbf{V}}_s$) &  & \multicolumn{2}{c}{0.043} & \multicolumn{2}{c}{0.049} & \multicolumn{2}{c}{0.041} & \multicolumn{2}{c}{0.045} \\
		GEE ($\widetilde{\mathbf{V}}_s$) & \multicolumn{1}{c}{$F$} & \multicolumn{2}{c}{0.040} & \multicolumn{2}{c}{0.046} & \multicolumn{2}{c}{\llap{$^\mathrm{c}$}\textbf{0.033}} & \multicolumn{2}{c}{0.038} \\
		NLMM$^{\mathrm{a}}$ &  & \multicolumn{2}{c}{0.053} & \multicolumn{2}{c}{\llap{$^\mathrm{c}$}\textbf{0.081}} & \multicolumn{2}{c}{\llap{$^\mathrm{c}$}\textbf{0.075}} & \multicolumn{2}{c}{\llap{$^\mathrm{c}$}\textbf{0.074}} \\
		\multicolumn{10}{l}{Scenario 3: ``misspecified'' in NLMM (Gamma)} \\
		GEE ($\widehat{\mathbf{V}}_s$) &  & 0.037 & \llap{$^\mathrm{c}$}\textbf{0.028} & 0.050 & 0.037 & 0.038 & 0.047 & 0.040 & 0.040 \\
		GEE ($\widetilde{\mathbf{V}}_s$) & \multicolumn{1}{c}{Wald} & 0.035 & \llap{$^\mathrm{c}$}\textbf{0.027} & 0.046 & \llap{$^\mathrm{c}$}\textbf{0.034} & 0.037 & 0.042 & \llap{$^\mathrm{c}$}\textbf{0.033} & \llap{$^\mathrm{c}$}\textbf{0.034} \\
		NLMM$^{\mathrm{a}}$ &  & 0.050 & 0.048 & 0.062 & 0.062 & 0.045 & 0.048 & 0.053 & \llap{$^\mathrm{c}$}\textbf{0.066} \\
		GEE ($\widehat{\mathbf{V}}_s$) &  & \multicolumn{2}{c}{0.035} & \multicolumn{2}{c}{0.056} & \multicolumn{2}{c}{0.045} & \multicolumn{2}{c}{0.053} \\
		GEE ($\widetilde{\mathbf{V}}_s$) & \multicolumn{1}{c}{$F$} & \multicolumn{2}{c}{\llap{$^\mathrm{c}$}\textbf{0.032}} & \multicolumn{2}{c}{0.051} & \multicolumn{2}{c}{0.043} & \multicolumn{2}{c}{0.046} \\
		NLMM$^{\mathrm{a}}$ &  & \multicolumn{2}{c}{0.052} & \multicolumn{2}{c}{\llap{$^\mathrm{c}$}\textbf{0.081}} & \multicolumn{2}{c}{0.064} & \multicolumn{2}{c}{\llap{$^\mathrm{c}$}\textbf{0.083}} \\ \hline
		
		\multicolumn{10}{c}{Bias} \\ \hline
		\multicolumn{10}{l}{Scenario 1: ``correctly specified'' in NLMM} \\
		\multicolumn{2}{l}{GEE ($\widehat{\mathbf{V}}_s$, $\widetilde{\mathbf{V}}_s$)$^{\mathrm{b}}$} & $<$0.001 & 0.001 & \llap{$<-$}0.001 & \llap{$-$}0.001 & \llap{$-$}0.007 & 0.008 & 0.007 & \llap{$-$}0.010 \\
		\multicolumn{2}{l}{NLMM$^{\mathrm{a}}$} & \llap{$-$}0.002 & \llap{$-$}0.003 & 0.001 & 0.001 & 0.002 & 0.012 & \llap{$-$}0.011 & \llap{$-$}0.008 \\
		\multicolumn{10}{l}{Scenario 2: ``misspecified'' in NLMM (Uniform)} \\
		\multicolumn{2}{l}{GEE ($\widehat{\mathbf{V}}_s$, $\widetilde{\mathbf{V}}_s$)$^{\mathrm{b}}$} & 0.002 & 0.002 & \llap{$<$}0.001 & \llap{$<$}0.001 & \llap{$-$}0.008 & \llap{$-$}0.011 & 0.002 & \llap{$-$}0.009 \\
		\multicolumn{2}{l}{NLMM$^{\mathrm{a}}$} & 0.001 & 0.003 & \llap{$<$}0.001 & \llap{$-$}0.002 & \llap{$-$}0.001 & \llap{$-$}0.008 & \llap{$-$}0.002 & \llap{$-$}0.008 \\
		\multicolumn{10}{l}{Scenario 3: ``misspecified'' in NLMM (Gamma)} \\
		\multicolumn{2}{l}{GEE ($\widehat{\mathbf{V}}_s$, $\widetilde{\mathbf{V}}_s$)$^{\mathrm{b}}$} & 0.001 & 0.002 & \llap{$<$}0.001 & \llap{$-$}0.001 & 0.002 & 0.002 & 0.002 & \llap{$-$}0.004 \\
		\multicolumn{2}{l}{NLMM$^{\mathrm{a}}$} & \llap{$-$}0.001 & 0.002 & \llap{$<$}0.001 & \llap{$-$}0.001 & 0.011 & \llap{$^{\mathrm{d}}$}\textbf{0.012} & \llap{$<$}0.001 & \llap{$-$}0.003 \\ \hline
		
		\multicolumn{10}{c}{MSE} \\ \hline
		\multicolumn{10}{l}{Scenario 1: ``correctly specified'' in NLMM} \\
		\multicolumn{2}{l}{GEE ($\widehat{\mathbf{V}}_s$, $\widetilde{\mathbf{V}}_s$)$^{\mathrm{b}}$} & 0.003 & 0.004 & 0.001 & 0.001 & 0.046 & 0.053 & 0.058 & 0.071 \\
		\multicolumn{2}{l}{NLMM$^{\mathrm{a}}$} & 0.003 & 0.004 & \llap{$<$}0.001 & 0.001 & 0.038 & 0.047 & 0.067 & 0.077 \\
		\multicolumn{10}{l}{Scenario 2: ``misspecified'' in NLMM (Uniform)} \\
		\multicolumn{2}{l}{GEE ($\widehat{\mathbf{V}}_s$, $\widetilde{\mathbf{V}}_s$)$^{\mathrm{b}}$} & 0.003 & 0.004 & 0.001 & 0.001 & 0.041 & 0.055 & 0.042 & 0.059 \\
		\multicolumn{2}{l}{NLMM$^{\mathrm{a}}$} & 0.003 & 0.004 & 0.001 & 0.001 & 0.040 & 0.049 & \llap{$^{\mathrm{d}}$}\textbf{0.062} & \llap{$^{\mathrm{d}}$}\textbf{0.078} \\
		\multicolumn{10}{l}{Scenario 3: ``misspecified'' in NLMM (Gamma)} \\
		\multicolumn{2}{l}{GEE ($\widehat{\mathbf{V}}_s$, $\widetilde{\mathbf{V}}_s$)$^{\mathrm{b}}$} & 0.003 & 0.004 & 0.001 & 0.001 & 0.039 & \llap{$^{\mathrm{d}}$}\textbf{0.055} & 0.019 & 0.027 \\
		\multicolumn{2}{l}{NLMM$^{\mathrm{a}}$} & 0.003 & 0.004 & 0.001 & 0.001 & 0.030 & 0.042 & 0.024 & 0.033 \\ \hline
		
		\multicolumn{10}{p{15.9cm}}{\scriptsize $^{\mathrm{a}}$Values of the NLMM have been calculated from simulations with a low convergence proportion for iterative calculation (see Subsection 3.3).} \\
		\multicolumn{10}{p{15.9cm}}{\scriptsize $^{\mathrm{b}}$Because GEE ($\widehat{\mathbf{V}}_s$) and GEE ($\widetilde{\mathbf{V}}_s$) differ only in variances, results of biases were combined as GEE ($\widehat{\mathbf{V}}_s$, $\widetilde{\mathbf{V}}_s$).} \\
		\multicolumn{10}{p{15.9cm}}{\scriptsize $^{\mathrm{c}}$Values that were $\leq 0.034$ or $\geq 0.066$ (binomial 99 \% upper confidential limit, $0.05\pm2.33\sqrt{0.05(1-0.05)/1000}$) are highlighted.} \\
		\multicolumn{10}{p{15.9cm}}{\scriptsize $^{\mathrm{d}}$Values for which there was more than a 0.01 inferior difference between GEE ($\widehat{\mathbf{V}}_s$, $\widetilde{\mathbf{V}}_s$) and the NLMM are highlighted.}\\
		\multicolumn{10}{p{15.9cm}}{\scriptsize The number of simulations was 1000.} \\

	\end{tabular}
	
\end{table}

Type-I error rates for the proposed Wald-type test and asymptotic $F$-test of GEE ($\widehat{\mathbf{V}}_s$) and GEE ($\widetilde{\mathbf{V}}_s$) were well controlled below the nominal level of 5\% for Scenarios 1--3 as shown in Tables 1 and 2.
However, the proposed Wald-type test and asymptotic $F$-test of GEE ($\widetilde{\mathbf{V}}_s$) were conservative.
In addition, type-I error rates for the proposed Wald-type test and asymptotic $F$-test of GEE ($\widehat{\mathbf{V}}_s$) were closer to the nominal level than those of GEE ($\widetilde{\mathbf{V}}_s$) despite the downward bias of $\widehat{\mathbf{V}}_s$ (as shown in Section 3.6).

In contrast, except in Scenario 3 (MAF = 0.25, $\beta_{K_{12}}$), type-I error rates for the Wald-type test of NLMM were inflated, as shown in Tables 1 and 2.
To be more specific, for Scenario 1 (MAF = 0.25), type-I error rates of NLMM were 0.067 ($H_0: \beta_{K_{21} Aa}=0$);
for Scenario 2 (MAF = 0.25), NLMM values were 0.075 ($H_0: \beta_{K_{el} Aa}=0$), 0.078 ($H_0: \beta_{K_{21} Aa}=0$), and 0.068 ($H_0: \beta_{K_{21} AA}=0$);
and for Scenario 3, the NLMM value was 0.068 ($H_0: \beta_{K_{el} Aa}=0$).
The results for MAF = 0.50 exhibit the same tendencies.
Because type-I error rates for Scenario 1 were closer to the nominal level than for Scenario 2, random-effects misspecification led to inflation of the type-I error rate in some instances.
Furthermore, except in Scenario 3 (MAF = 0.25, $\beta_{K_{12}}$), type-I error rates for the asymptotic $F$-test of NLMM were inflated, as shown in Tables 1 and 2.

Except in Scenario 2 (MAF = 0.25) and Scenario 3, the biases did not differ greatly between GEE ($\widehat{\mathbf{V}}_s$, $\widetilde{\mathbf{V}}_s$) and NLMM, as shown in Tables 1 and 2.
Therefore, the biases of GEE ($\widehat{\mathbf{V}}_s$, $\widetilde{\mathbf{V}}_s$) might be slightly larger than those of the NLMM.

The MSEs did not differ greatly between GEE ($\widehat{\mathbf{V}}_s$, $\widetilde{\mathbf{V}}_s$) and NLMM, except in  $\beta_{K_{12} AA}$, $\beta_{K_{21} Aa}$ and  $\beta_{K_{21} AA}$, as shown in Tables 1 and 2.
For  $\beta_{K_{12} AA}$, MSEs of GEE ($\widehat{\mathbf{V}}_s$, $\widetilde{\mathbf{V}}_s$) were larger than those of NLMM.
For $\beta_{K_{21} Aa}$ and  $\beta_{K_{21} AA}$, MSEs of the NLMM were larger than those of GEE ($\widehat{\mathbf{V}}_s$, $\widetilde{\mathbf{V}}_s$).
Therefore, the MSEs of GEE ($\widehat{\mathbf{V}}_s$, $\widetilde{\mathbf{V}}_s$) might not differ significantly from those of the NLMM.

In summary, the proposed Wald-type test and asymptotic $F$-test of GEE ($\widehat{\mathbf{V}}_s$) and GEE ($\widetilde{\mathbf{V}}_s$) could control type-I error rates in Scenarios 1--3, and the type-I error rates of the Wald-type test and asymptotic $F$-test of the NLMM were inflated in some cases.

\subsection{Powers}
In this Section, we compare the powers of GEE ($\widehat{\mathbf{V}}_s$), GEE ($\widetilde{\mathbf{V}}_s$), and NLMM in the four following scenarios:
\begin{description}
	\item \textbf{Scenario 4} 
	The coefficients $\beta_{V_d Aa}$ and $\beta_{V_d AA}$ were set to $0.05 \beta_{V_d}$, and the other effect parameters were set to 0.
	In Scenario 4, $\beta_{V_d Aa}=0.05 \beta_{V_d}$ implies that $\log V_d^{(i)}$ of the population with SNP genotype ``Aa'' increases by 5\% compared with the population with SNP genotype ``aa''.
	\item \textbf{Scenario 5} 
	The coefficients $\beta_{K_{el} Aa}$ and $\beta_{K_{el} AA}$ were set to $0.05 \beta_{K_{el}}$, and the other effect parameters were set to 0.
	\item \textbf{Scenario 6} 
	The coefficients $\beta_{K_{12} Aa}$ and $\beta_{K_{12} AA}$ were set to $0.30 \beta_{K_{12}}$, and the other effect parameters were set to 0.
	\item \textbf{Scenario 7} 
	The coefficients $\beta_{K_{21} Aa}$ and $\beta_{K_{21} AA}$ were set to $0.50 \beta_{K_{21}}$, and the other effect parameters were set to 0.
\end{description}

In these scenarios, the random-effects distributions were set to ``correctly specified'' in NLMM.
We assumed that the random-effect vector $\boldsymbol{\gamma}_i$ is normally distributed with a mean vector $\mathbf{0}$ and a covariance matrix $\mathrm{diag}(\tau_{V_d}^2, \tau_{K_{12}}^2, \tau_{K_{21}}^2)$, where the standard deviation for random effects was set to 0.12, 0.68, and 0.89, respectively.

Powers, biases, and MSEs of GEE ($\widehat{\mathbf{V}}_s$), GEE ($\widetilde{\mathbf{V}}_s$), and NLMM are shown in Tables 3 and 4.
Because GEE ($\widehat{\mathbf{V}}_s$) and GEE ($\widetilde{\mathbf{V}}_s$) differ only in variances, results of biases and MSEs were combined as GEE ($\widehat{\mathbf{V}}_s$), GEE ($\widetilde{\mathbf{V}}_s$) in Tables 3 and 4.

\begin{table}
	\small
	\caption{Powers, biases, and MSEs of GEE ($\widehat{\mathbf{V}}_s$), GEE ($\widetilde{\mathbf{V}}_s$), and NLMM for MAF = 0.25 in Scenarios~4--7.}
	\label{t:power}
	\centering
	\begin{tabular}{lccccccccc}\hline
		&  & \multicolumn{2}{c}{Scenario 4} & \multicolumn{2}{c}{Scenario 5} & \multicolumn{2}{c}{Scenario 6} & \multicolumn{2}{c}{Scenario 7} \\
		Parameter &  & $\beta_{V_dAa}$ & $\beta_{V_dAA}$ & $\beta_{K_{el}Aa}$ & $\beta_{K_{el}AA}$ & $\beta_{K_{12}Aa}$ & $\beta_{K_{12}AA}$ & $\beta_{K_{21}Aa}$ & $\beta_{K_{21}AA}$ \\ \hline
		
		\multicolumn{10}{c}{Power} \\ \hline
		GEE ($\widehat{\mathbf{V}}_s$) &  & \multicolumn{1}{r}{0.765} & \multicolumn{1}{r}{\llap{$^\mathrm{c}$}\textbf{0.272}} & \multicolumn{1}{r}{0.897} & \multicolumn{1}{r}{\llap{$^\mathrm{c}$}\textbf{0.294}} & \multicolumn{1}{r}{0.819} & \multicolumn{1}{r}{\llap{$^\mathrm{c}$}\textbf{0.186}} & \multicolumn{1}{r}{0.086} & \multicolumn{1}{r}{0.069} \\
		GEE ($\widetilde{\mathbf{V}}_s$) & Wald & \multicolumn{1}{r}{0.762} & \multicolumn{1}{r}{\llap{$^\mathrm{c}$}\textbf{0.246}} & \multicolumn{1}{r}{0.894} & \multicolumn{1}{r}{\llap{$^\mathrm{c}$}\textbf{0.263}} & \multicolumn{1}{r}{0.811} & \multicolumn{1}{r}{\llap{$^\mathrm{c}$}\textbf{0.142}} & \multicolumn{1}{r}{0.075} & \multicolumn{1}{r}{0.047} \\
		NLMM$^{\mathrm{a}}$ &  & \multicolumn{1}{r}{0.786} & \multicolumn{1}{r}{0.418} & \multicolumn{1}{r}{0.957} & \multicolumn{1}{r}{0.484} & \multicolumn{1}{r}{0.919} & \multicolumn{1}{r}{0.481} & \multicolumn{1}{r}{0.152} & \multicolumn{1}{r}{0.100} \\
		GEE ($\widehat{\mathbf{V}}_s$) &  & \multicolumn{2}{c}{\llap{$^\mathrm{c}$}\textbf{0.660}} & \multicolumn{2}{c}{\llap{$^\mathrm{c}$}\textbf{0.757}} & \multicolumn{2}{c}{\llap{$^\mathrm{c}$}\textbf{0.610}} & \multicolumn{2}{c}{\llap{$^\mathrm{c}$}\textbf{0.072}} \\
		GEE ($\widetilde{\mathbf{V}}_s$) & $F$ & \multicolumn{2}{c}{\llap{$^\mathrm{c}$}\textbf{0.638}} & \multicolumn{2}{c}{\llap{$^\mathrm{c}$}\textbf{0.728}} & \multicolumn{2}{c}{\llap{$^\mathrm{c}$}\textbf{0.551}} & \multicolumn{2}{c}{\llap{$^\mathrm{c}$}\textbf{0.055}} \\
		NLMM$^{\mathrm{a}}$ &  & \multicolumn{2}{c}{0.777} & \multicolumn{2}{c}{0.948} & \multicolumn{2}{c}{0.934} & \multicolumn{2}{c}{0.179} \\ \hline
		
		\multicolumn{10}{c}{Bias} \\ \hline
		GEE ($\widehat{\mathbf{V}}_s$, $\widetilde{\mathbf{V}}_s$)$^{\mathrm{b}}$ &  & \multicolumn{1}{r}{\llap{$-$}0.040} & \multicolumn{1}{r}{\llap{$-$}0.041} & \multicolumn{1}{r}{\llap{$-$}0.001} & \multicolumn{1}{r}{\llap{$-$}0.001} & \multicolumn{1}{r}{\llap{$^{\mathrm{d}}$}\textbf{0.021}} & \multicolumn{1}{r}{\llap{$^{\mathrm{d}}$}\textbf{0.048}} & \multicolumn{1}{r}{\llap{$-$}0.014} & \multicolumn{1}{r}{\llap{$^{\mathrm{d}}$}\textbf{0.040}} \\
		NLMM$^{\mathrm{a}}$ &  & \multicolumn{1}{r}{\llap{$-$}0.039} & \multicolumn{1}{r}{\llap{$-$}0.037} & \multicolumn{1}{r}{\llap{$-$}0.002} & \multicolumn{1}{r}{\llap{$-$}0.002} & \multicolumn{1}{r}{\llap{$-$}0.007} & \multicolumn{1}{r}{\llap{$-$}0.016} & \multicolumn{1}{r}{\llap{$-$}0.010} & \multicolumn{1}{r}{0.017} \\ \hline
		
		\multicolumn{10}{c}{MSE} \\ \hline
		GEE ($\widehat{\mathbf{V}}_s$, $\widetilde{\mathbf{V}}_s$)$^{\mathrm{b}}$ &  & \multicolumn{1}{r}{0.010} & \multicolumn{1}{r}{0.017} & \multicolumn{1}{r}{\llap{$<$}0.001} & \multicolumn{1}{r}{0.001} & \multicolumn{1}{r}{0.034} & \multicolumn{1}{r}{\llap{$^{\mathrm{d}}$}\textbf{0.129}} & \multicolumn{1}{r}{0.042} & \multicolumn{1}{r}{0.164} \\
		NLMM$^{\mathrm{a}}$ &  & \multicolumn{1}{r}{0.010} & \multicolumn{1}{r}{0.016} & \multicolumn{1}{r}{\llap{$<$}0.001} & \multicolumn{1}{r}{0.001} & \multicolumn{1}{r}{0.028} & \multicolumn{1}{r}{0.104} & \multicolumn{1}{r}{0.050} & \multicolumn{1}{r}{0.172} \\ \hline
		
		\multicolumn{10}{p{15.9cm}}{\footnotesize $^{\mathrm{a}}$Values of the NLMM have been calculated from simulations with a low convergence proportion for iterative calculation (see Subsection 3.3).} \\
		\multicolumn{10}{p{15.9cm}}{\footnotesize $^{\mathrm{b}}$Because GEE ($\widehat{\mathbf{V}}_s$) and GEE ($\widetilde{\mathbf{V}}_s$) differ only in variances, results of biases were combined as GEE ($\widehat{\mathbf{V}}_s$, $\widetilde{\mathbf{V}}_s$).} \\
		\multicolumn{10}{p{15.9cm}}{\footnotesize $^{\mathrm{c}}$Values for which there was more than a 0.1 inferior difference between GEE ($\widehat{\mathbf{V}}_s$), GEE ($\widetilde{\mathbf{V}}_s$), and NLMM are highlighted.} \\
		\multicolumn{10}{p{15.9cm}}{\footnotesize $^{\mathrm{d}}$Values for which there was more than a 0.01 inferior difference between GEE ($\widehat{\mathbf{V}}_s$, $\widetilde{\mathbf{V}}_s$) and the NLMM are highlighted.}\\
		\multicolumn{10}{p{15.9cm}}{\footnotesize The number of simulations was 1000.} \\
		
	\end{tabular}
\end{table}

\begin{table}
	\small
	\caption{Powers, biases, and MSEs of GEE ($\widehat{\mathbf{V}}_s$), GEE ($\widetilde{\mathbf{V}}_s$), and NLMM for MAF = 0.50 in Scenarios~4--7.}
	\label{t:power050}
	\centering
	\begin{tabular}{lccccccccc} \hline
		&  & \multicolumn{2}{c}{Scenario 4} & \multicolumn{2}{c}{Scenario 5} & \multicolumn{2}{c}{Scenario 6} & \multicolumn{2}{c}{Scenario 7} \\
		Parameter &  & $\beta_{V_dAa}$ & $\beta_{V_dAA}$ & $\beta_{K_{el}Aa}$ & $\beta_{K_{el}AA}$ & $\beta_{K_{12}Aa}$ & $\beta_{K_{12}AA}$ & $\beta_{K_{21}Aa}$ & $\beta_{K_{21}AA}$ \\ \hline
		
		\multicolumn{10}{c}{Power} \\ \hline
		GEE ($\widehat{\mathbf{V}}_s$) &  & \multicolumn{1}{r}{0.875} & \multicolumn{1}{r}{0.767} & \multicolumn{1}{r}{0.810} & \multicolumn{1}{r}{\llap{$^{\mathrm{c}}$}\textbf{0.687}} & \multicolumn{1}{r}{\llap{$^{\mathrm{c}}$}\textbf{0.696}} & \multicolumn{1}{r}{\llap{$^{\mathrm{c}}$}\textbf{0.548}} & \multicolumn{1}{r}{0.065} & \multicolumn{1}{r}{0.061} \\
		GEE ($\widetilde{\mathbf{V}}_s$) & Wald & \multicolumn{1}{r}{0.868} & \multicolumn{1}{r}{0.752} & \multicolumn{1}{r}{0.806} & \multicolumn{1}{r}{\llap{$^{\mathrm{c}}$}\textbf{0.665}} & \multicolumn{1}{r}{\llap{$^{\mathrm{c}}$}\textbf{0.679}} & \multicolumn{1}{r}{\llap{$^{\mathrm{c}}$}\textbf{0.532}} & \multicolumn{1}{r}{0.055} & \multicolumn{1}{r}{0.054} \\
		NLMM$^{\mathrm{a}}$ &  & \multicolumn{1}{r}{0.918} & \multicolumn{1}{r}{0.815} & \multicolumn{1}{r}{0.889} & \multicolumn{1}{r}{0.805} & \multicolumn{1}{r}{0.845} & \multicolumn{1}{r}{0.738} & \multicolumn{1}{r}{0.150} & \multicolumn{1}{r}{0.126} \\
		GEE ($\widehat{\mathbf{V}}_s$) &  & \multicolumn{2}{c}{0.914} & \multicolumn{2}{c}{0.860} & \multicolumn{2}{c}{\llap{$^{\mathrm{c}}$}\textbf{0.747}} & \multicolumn{2}{c}{0.081} \\
		GEE ($\widetilde{\mathbf{V}}_s$) & $F$ & \multicolumn{2}{c}{0.911} & \multicolumn{2}{c}{0.848} & \multicolumn{2}{c}{\llap{$^{\mathrm{c}}$}\textbf{0.732}} & \multicolumn{2}{c}{\llap{$^{\mathrm{c}}$}\textbf{0.066}} \\
		NLMM$^{\mathrm{a}}$ &  & \multicolumn{2}{c}{0.944} & \multicolumn{2}{c}{0.934} & \multicolumn{2}{c}{0.903} & \multicolumn{2}{c}{0.177} \\ \hline
		
		\multicolumn{10}{c}{Bias} \\ \hline
		GEE ($\widehat{\mathbf{V}}_s$, $\widetilde{\mathbf{V}}_s$)$^{\mathrm{b}}$ &  & \multicolumn{1}{r}{0.004} & \multicolumn{1}{r}{0.005} & \multicolumn{1}{r}{\llap{$-$}0.001} & \multicolumn{1}{r}{\llap{$-$}0.001} & \multicolumn{1}{r}{0.017} & \multicolumn{1}{r}{0.021} & \multicolumn{1}{r}{0.017} & \multicolumn{1}{r}{0.009} \\
		NLMM$^{\mathrm{a}}$ &  & \multicolumn{1}{r}{0.003} & \multicolumn{1}{r}{0.004} & \multicolumn{1}{r}{\llap{$-$}0.001} & \multicolumn{1}{r}{$<$0.001} & \multicolumn{1}{r}{\llap{$-$}0.011} & \multicolumn{1}{r}{\llap{$-$}0.014} & \multicolumn{1}{r}{\llap{$-$}0.008} & \multicolumn{1}{r}{\llap{$-$}0.007} \\ \hline
		
		\multicolumn{10}{c}{MSE} \\ \hline
		GEE ($\widehat{\mathbf{V}}_s$, $\widetilde{\mathbf{V}}_s$)$^{\mathrm{b}}$ &  & \multicolumn{1}{r}{0.004} & \multicolumn{1}{r}{0.005} & \multicolumn{1}{r}{0.001} & \multicolumn{1}{r}{0.001} & \multicolumn{1}{r}{0.042} & \multicolumn{1}{r}{0.059} & \multicolumn{1}{r}{0.056} & \multicolumn{1}{r}{0.075} \\
		NLMM$^{\mathrm{a}}$ &  & \multicolumn{1}{r}{0.003} & \multicolumn{1}{r}{0.004} & \multicolumn{1}{r}{0.001} & \multicolumn{1}{r}{0.001} & \multicolumn{1}{r}{0.034} & \multicolumn{1}{r}{0.049} & \multicolumn{1}{r}{0.060} & \multicolumn{1}{r}{0.080} \\ \hline
		
		\multicolumn{10}{p{15.9cm}}{\footnotesize $^{\mathrm{a}}$Values of the NLMM have been calculated from simulations with a low convergence proportion for iterative calculation (see Subsection 3.3).} \\
		\multicolumn{10}{p{15.9cm}}{\footnotesize $^{\mathrm{b}}$Because GEE ($\widehat{\mathbf{V}}_s$) and GEE ($\widetilde{\mathbf{V}}_s$) differ only in variances, results of biases were combined as GEE ($\widehat{\mathbf{V}}_s$, $\widetilde{\mathbf{V}}_s$).} \\
		\multicolumn{10}{p{15.9cm}}{\footnotesize $^{\mathrm{c}}$Values for which there was more than a 0.1 inferior difference between GEE ($\widehat{\mathbf{V}}_s$), GEE ($\widetilde{\mathbf{V}}_s$), and NLMM are highlighted.} \\
		\multicolumn{10}{p{15.9cm}}{\footnotesize The number of simulations was 1000.} \\
		
	\end{tabular}
\end{table}

NLMM was the most powerful method.
The power of the Wald-type test of GEE ($\widehat{\mathbf{V}}_s$) was about 0.021--0.295 smaller, and that of GEE ($\widetilde{\mathbf{V}}_s$) was about 0.024--0.339 smaller, than that of NLMM (Table 3).
The power of the asymptotic $F$-test of GEE ($\widehat{\mathbf{V}}_s$) was about 0.030--0.324 smaller, and that of GEE ($\widetilde{\mathbf{V}}_s$) was about 0.033--0.383 smaller, than that of NLMM (Table 3).
Furthermore, because $\widehat{\mathbf{V}}_s<\widetilde{\mathbf{V}}_s$, the power of GEE ($\widehat{\mathbf{V}}_s$) was higher than that of GEE ($\widetilde{\mathbf{V}}_s$) in all scenarios.
Power increased with the sum of subgroup sample size of $n_{aa}+n_{Aa}$ or $n_{aa}+n_{AA}$.

The biases and MSEs of GEE ($\widehat{\mathbf{V}}_s$, $\widetilde{\mathbf{V}}_s$) were larger than those of NLMM, as shown in Tables 3 and 4.
However, these results did not indicate a discernible trend.

Therefore, unless influenced by inflation of the type-I error rates, the power of NLMM might be greater than those of GEE ($\widehat{\mathbf{V}}_s$) and GEE ($\widetilde{\mathbf{V}}_s$).

\subsection{Computation time and convergence proportion for iterative calculation}
In this Section, we compare computation time and convergence proportion for iterative calculation of GEE ($\widehat{\mathbf{V}}_s$), GEE ($\widetilde{\mathbf{V}}_s$), and NLMM.
All computations were performed on a personal computer with a 3.0-GHz Intel Core 2 Duo CPU and 3.25 GB of RAM running 32-bit Windows XP.
Computation times and convergence proportions for iterative calculations for each scenario are shown in Table 5.

\begin{table}
	\small
	\caption{Computation time per 1000 SNPs, ratio of GEE ($\widehat{\mathbf{V}}_s$) computation time, and convergence proportion for iterative calculations}
	\label{t:comptime}
	\centering
	\begin{tabular}{lrrrrrr} \hline
		& \multicolumn{3}{c}{MAF = 0.25} & \multicolumn{3}{c}{MAF = 0.50} \\ \cline{2-7}
		& \multicolumn{1}{c}{Computation} & \multicolumn{1}{c}{} & \multicolumn{1}{c}{} & \multicolumn{1}{c}{Computation} & \multicolumn{1}{c}{} & \multicolumn{1}{c}{} \\
		& \multicolumn{1}{c}{time per} & \multicolumn{1}{c}{Ratio of} & \multicolumn{1}{c}{Percent} & \multicolumn{1}{c}{time per} & \multicolumn{1}{c}{Ratio of} & \multicolumn{1}{c}{Percent} \\
		& \multicolumn{1}{c}{1000 SNPs} & \multicolumn{1}{c}{GEE ($\widehat{\mathbf{V}}_s$)} & \multicolumn{1}{c}{convergence} & \multicolumn{1}{c}{1000 SNPs} & \multicolumn{1}{c}{GEE ($\widehat{\mathbf{V}}_s$)} & \multicolumn{1}{c}{convergence} \\
		& \multicolumn{1}{c}{(minutes)} & \multicolumn{1}{c}{} & \multicolumn{1}{c}{} & \multicolumn{1}{c}{(minutes)} & \multicolumn{1}{c}{} & \multicolumn{1}{c}{} \\ \hline
		
		Scenario 1 &  &  &  &  &  &  \\
		GEE ($\widehat{\mathbf{V}}_s$) & 8.7 & 1.0 & 100.0 & 4.9 & 1.0 & 100.0 \\
		GEE ($\widetilde{\mathbf{V}}_s$) & 10.1 & 1.2 & 100.0 & 6.5 & 1.3 & 100.0 \\
		NLMM & 8050.8 & 928.7 & 65.0 & 7184.3 & 1459.0 & 61.6 \\
		Scenario 2 &  &  &  &  &  &  \\
		GEE ($\widehat{\mathbf{V}}_s$) & 7.7 & 1.0 & 100.0 & 8.3 & 1.0 & 100.0 \\
		GEE ($\widetilde{\mathbf{V}}_s$) & 9.0 & 1.2 & 100.0 & 9.6 & 1.2 & 100.0 \\
		NLMM & 6709.7 & 870.7 & 63.7 & 8066.3 & 977.1 & 66.5 \\
		Scenario 3 &  &  &  &  &  &  \\
		GEE ($\widehat{\mathbf{V}}_s$) & 8.0 & 1.0 & 100.0 & 4.2 & 1.0 & 100.0 \\
		GEE ($\widetilde{\mathbf{V}}_s$) & 9.3 & 1.2 & 100.0 & 5.5 & 1.3 & 100.0 \\
		NLMM & 6753.9 & 849.3 & 57.7 & 4679.5 & 1116.7 & 58.1 \\
		Scenario 4 &  &  &  &  &  &  \\
		GEE ($\widehat{\mathbf{V}}_s$) & 7.6 & 1.0 & 100.0 & 4.2 & 1.0 & 100.0 \\
		GEE ($\widetilde{\mathbf{V}}_s$) & 8.9 & 1.2 & 100.0 & 5.8 & 1.4 & 100.0 \\
		NLMM & 7807.9 & 1023.6 & 64.5 & 6804.6 & 1630.3 & 66.1 \\
		Scenario 5 &  &  &  &  &  &  \\
		GEE ($\widehat{\mathbf{V}}_s$) & 4.7 & 1.0 & 100.0 & 7.2 & 1.0 & 100.0 \\
		GEE ($\widetilde{\mathbf{V}}_s$) & 5.9 & 1.3 & 100.0 & 8.6 & 1.2 & 100.0 \\
		NLMM & 6169.9 & 1323.1 & 66.0 & 8128.7 & 1136.7 & 65.4 \\
		Scenario 6 &  &  &  &  &  &  \\
		GEE ($\widehat{\mathbf{V}}_s$) & 7.5 & 1.0 & 100.0 & 7.6 & 1.0 & 100.0 \\
		GEE ($\widetilde{\mathbf{V}}_s$) & 9.2 & 1.2 & 100.0 & 8.9 & 1.2 & 100.0 \\
		NLMM & 7779.4 & 1033.0 & 62.3 & 10864.2 & 1422.5 & 62.0 \\
		Scenario 7 &  &  &  &  &  &  \\
		GEE ($\widehat{\mathbf{V}}_s$) & 8.4 & 1.0 & 100.0 & 7.9 & 1.0 & 100.0 \\
		GEE ($\widetilde{\mathbf{V}}_s$) & 10.3 & 1.2 & 100.0 & 9.7 & 1.2 & 100.0 \\
		NLMM & 10988.9 & 1303.1 & 62.6 & 10168.4 & 1279.4 & 57.1 \\ \hline
	\end{tabular}
\end{table}

GEE ($\widehat{\mathbf{V}}_s$) was the fastest method, GEE ($\widetilde{\mathbf{V}}_s$) required 1.2--1.4-fold longer, and NLMM required 849.3--1630.3-fold longer than GEE ($\widehat{\mathbf{V}}_s$), as shown in Table 5.
For instance, computational times of GEE ($\widehat{\mathbf{V}}_s$), GEE ($\widetilde{\mathbf{V}}_s$), and NLMM in Scenario 1 (MAF = 0.25), whose dataset includes 1000 SNPs, were 8.7, 10.1, and 8050.8 minutes, respectively.

The convergence success of GEE ($\widehat{\mathbf{V}}_s$) and GEE ($\widetilde{\mathbf{V}}_s$) was achieved perfectly in all scenarios.
However, the convergence success of NLMM did not even reach 70{\%};
NLMM tended not to converge for data sets with relatively large random effects.
In genome-wide PGx studies, oligonucleotide SNP arrays can provide information about 100,000--4,300,000 SNPs.
For instance, if the convergence success is 70{\%} with 100,000 SNPs, we derived no information for 30,000 SNPs.
Therefore, the simulation results suggested that GEE ($\widehat{\mathbf{V}}_s$) and GEE ($\widetilde{\mathbf{V}}_s$) perform at a relatively high speed with stable computation in genome-wide settings.

\section{Application to an actual genome-wide PGx study data}
We determined GEE ($\widehat{\mathbf{V}}_s$) and GEE ($\widetilde{\mathbf{V}}_s$) using a genome-wide PGx study \cite{Sato}, which analyzed plasma concentrations of gemcitabine ($n = 233$ patients) with respect to 109,365 gene-centric SNPs using the Sentrix Human-1 Genotyping BeadChip (Illumina Inc., San Diego, CA).
For reducing false positives, the $P$-value cutoff of asymptotic $F$-tests was set to $\alpha=1.14 \times 10^{-7}=0.05/109365/4$ as a simple Bonferroni adjustment.
The results showed that 82 SNPs were significant by GEE ($\widehat{\mathbf{V}}_s$), and 79 SNPs were significant by GEE ($\widetilde{\mathbf{V}}_s$).
These computations of the 109,365 SNPs were finished in 16.3 hours for GEE ($\widehat{\mathbf{V}}_s$) and 19.9 hours for GEE ($\widetilde{\mathbf{V}}_s$).

\begin{table}
	\small
	\caption{Comparison of SNP analyses using GEE ($\widehat{\mathbf{V}}_s$) and NLMM}
	\label{t:application}
	\centering
	\begin{tabular}{llrrrrcc} \hline
		\multicolumn{1}{c}{Method} & \multicolumn{1}{c}{Parameter} & \multicolumn{1}{c}{Estimate} & \multicolumn{1}{c}{S.E.} & \multicolumn{1}{c}{d.f.} & \multicolumn{1}{c}{$P$-value} & denom. d.f. & $P$-value \\
		\multicolumn{1}{c}{} & \multicolumn{1}{c}{} & \multicolumn{1}{c}{} & \multicolumn{1}{c}{} & \multicolumn{1}{c}{(Wald)} & \multicolumn{1}{c}{(Wald)} & ($F$) & ($F$) \\ \hline
		
		GEE ($\widehat{\mathbf{V}}_s$) & $\beta_{V_d}$ & 3.87 & 0.029 & 44.0 & 3.55$\times10^{-59}$ & -- & -- \\
		& $\beta_{V_dAa}$ & \llap{$-$}0.03 & 0.057 & 24.4 & 5.69$\times10^{-01}$ & \multicolumn{1}{r}{3.0} & \multicolumn{1}{r}{2.35$\times10^{-01}$} \\
		& $\beta_{V_dAA}$ & 0.18 & 0.085 & 2.5 & 1.40$\times10^{-01}$ &  &  \\
		& $\beta_{K_{el}}$ & 1.31 & 0.017 & 44.4 & 1.34$\times10^{-48}$ & -- & -- \\
		& $\beta_{K_{el}Aa}$ & \llap{$-$}0.03 & 0.036 & 24.2 & 4.40$\times10^{-01}$ & \multicolumn{1}{r}{10.4} & \multicolumn{1}{r}{\llap{$^{\mathrm{a}}$}$\mathbf{6.74\times10^{-10}}$} \\
		& $\beta_{K_{el}AA}$ & \llap{$-$}0.87 & 0.036 & 7.2 & 3.62$\times10^{-08}$ &  &  \\
		& $\beta_{K_{12}}$ & \llap{$-$}2.30 & 0.066 & 16.2 & 1.27$\times10^{-16}$ & -- & -- \\
		& $\beta_{K_{12}Aa}$ & \llap{$-$}0.10 & 0.124 & 41.6 & 4.42$\times10^{-01}$ & \multicolumn{1}{r}{24.0} & \multicolumn{1}{r}{\llap{$^{\mathrm{a}}$}$\mathbf{<1.11\times10^{-16}}$} \\
		& $\beta_{K_{12}AA}$ & 2.85 & 0.079 & 17.2 & 1.15$\times10^{-17}$ &  &  \\
		& $\beta_{K_{21}}$ & \llap{$-$}0.80 & 0.087 & 14.0 & 2.64$\times10^{-07}$ & -- & -- \\
		& $\beta_{K_{21}Aa}$ & 0.04 & 0.160 & 38.9 & 7.99$\times10^{-01}$ & \multicolumn{1}{r}{11.5} & \multicolumn{1}{r}{\llap{$^{\mathrm{a}}$}$\mathbf{1.15\times10^{-11}}$} \\
		& $\beta_{K_{21}AA}$ & \llap{$-$}4.73 & 0.157 & 7.4 & 4.69$\times10^{-09}$ &  &  \\ \hline
		NLMM & $\beta_{V_d}$ & 3.71 & 0.036 & 231.0 & 6.74$\times10^{-194}$ & -- & -- \\
		& $\beta_{V_dAa}$ & \multicolumn{1}{c}{--} & \multicolumn{1}{c}{--} & \multicolumn{1}{c}{--} & \multicolumn{1}{c}{--} & -- & -- \\
		& $\beta_{V_dAA}$ & 0.33 & 0.089 & 231.0 & 2.83$\times10^{-04}$ & \multicolumn{1}{r}{231.0} & \multicolumn{1}{r}{2.83$\times10^{-04}$} \\
		& $\beta_{K_{el}}$ & 1.39 & 0.022 & 231.0 & 8.42$\times10^{-149}$ & -- & -- \\
		& $\beta_{K_{el}Aa}$ & \multicolumn{1}{c}{--} & \multicolumn{1}{c}{--} & \multicolumn{1}{c}{--} & \multicolumn{1}{c}{--} & -- & -- \\
		& $\beta_{K_{el}AA}$ & \llap{$-$}0.37 & 0.097 & 231.0 & 1.98$\times10^{-04}$ & \multicolumn{1}{r}{231.0} & \multicolumn{1}{r}{1.98$\times10^{-04}$} \\
		& $\beta_{K_{12}}$ & \llap{$-$}2.12 & 0.085 & 231.0 & 4.02$\times10^{-67}$ & -- & -- \\
		& $\beta_{K_{12}Aa}$ & \multicolumn{1}{c}{--} & \multicolumn{1}{c}{--} & \multicolumn{1}{c}{--} & \multicolumn{1}{c}{--} & -- & -- \\
		& $\beta_{K_{12}AA}$ & 1.40 & 0.431 & 231.0 & 1.32$\times10^{-03}$ & \multicolumn{1}{r}{231.0} & \multicolumn{1}{r}{1.32$\times10^{-03}$} \\
		& $\beta_{K_{21}}$ & \llap{$-$}0.52 & 0.071 & 231.0 & 2.85$\times10^{-12}$ & -- & -- \\
		& $\beta_{K_{21}Aa}$ & \multicolumn{1}{c}{--} & \multicolumn{1}{c}{--} & \multicolumn{1}{c}{--} & \multicolumn{1}{c}{--} & -- & -- \\
		& $\beta_{K_{21}AA}$ & \llap{$-$}3.72 & 0.532 & 231.0 & 2.80$\times10^{-11}$ & \multicolumn{1}{r}{231.0} & \multicolumn{1}{r}{\llap{$^{\mathrm{a}}$}$\mathbf{2.80\times10^{-11}}$} \\ \hline
		
		\multicolumn{8}{p{15.5cm}}{\footnotesize $^{\mathrm{a}}$Values that the coefficients of a SNP effects were smaller than the significance level $\alpha=1.14\times10^{-7}$ are highlighted.} \\
		\multicolumn{8}{p{15.5cm}}{\footnotesize Abbreviations: S.E. = standard error, d.f. = degrees of freedom, denom. = denominator.}
		
	\end{tabular}
\end{table}

Table 6 shows a result for a SNP (rs234630) chosen from among the 82 SNPs significant by GEE ($\widehat{\mathbf{V}}_s$).
Note that the tests for the null hypotheses indicated that a SNP does not affect PK parameters (e.g., $H_0: \beta_{V_d Aa}=\beta_{V_d AA}=0$).
Further, results of NLMM with a Gaussian random-effects vector $\boldsymbol{\gamma}_i \sim \mathrm{N}(\mathbf{0}, \mathrm{diag}(\tau_{K_{12}}^2, \tau_{K_{21}}^2))$ after variable selection are also summarized in Table 6, because NLMM with a Gaussian random-effects vector $\boldsymbol{\gamma}_i \sim \mathrm{N}(\mathbf{0}, \mathrm{diag}(\tau_{V_d}^2, \tau_{K_{12}}^2, \tau_{K_{21}}^2))$ failed to converge.
The convergence success of NLMM without variable selection was 54.9 $\%$ (45 SNPs) among the 82 SNPs.
Computations for NLMM of the 82 SNPs were finished in 24.1 hours.

According to the results obtained based on GEE ($\widehat{\mathbf{V}}_s$), the coefficients of the SNP effects, $(\beta_{K_{el} Aa}, \beta_{K_{el} AA})$, $(\beta_{K_{12} Aa}, \beta_{K_{12} AA})$, and $(\beta_{K_{21} Aa}, \beta_{K_{21} AA})$ were statistically significant at $\alpha$.
In contrast, NLMM-based results indicated that only the coefficients $(\beta_{K_{21} Aa}, \beta_{K_{21} AA)}$ were statistically significant.
As a result, both GEE ($\widehat{\mathbf{V}}_s$) and NLMM indicated that the SNP affected PK parameter $K_{21}$ of gemcitabine.

In addition, we conducted similar analyses for all 82 significant SNPs, and observed that 74 of them (90.2{\%}) were statistically significant at $\alpha$ by both GEE ($\widehat{\mathbf{V}}_s$) and NLMM.

Therefore, we consider that GEE ($\widehat{\mathbf{V}}_s$) is a suitable alternative method for analyzing population PK data in genome-wide PGx studies.

\section{Discussion}
NLMM, which accounts for inter-individual variability in PK parameters, is a useful method for analyzing a genetic polymorphism in relation to population PK data \cite{Lindstrom,Davidian1993,Davidian1995,Vonesh,Wolfinger}.
However, when applying an NLMM to large-scale data, three problems occur in association with an assumption of random effects:
(i) computation time \cite{Davidian1993};
(ii) convergence of iterative calculation \cite{Pinheiro,Zhang};
and (iii) random-effects misspecification \cite{Hartford}.
In fact, the results of simulations show that NLMM was the slowest and the most computationally unstable;
furthermore, the type-I error rate of NLMM was inflated in some cases of random-effects misspecification.
As an alternative effective approach to resolving these issues, in this article we proposed valid inference methods for using GEE even under inter-individual variability, and provided theoretical justifications of the proposed GEE estimators for population PK data.
The proposed GEE methods applied a potentially misspecified model \cite{White,Yi} to account for inter-individual variability in PK parameters.
The proposed GEE estimator, $\widehat{\boldsymbol{\beta}}$, can be interpreted as the population-weighted average of the individual parameter vector, $\boldsymbol{\beta}_i$, under the true model.
The effectiveness of the proposed method was demonstrated through simulations and an application to a genome-wide PGx study \cite{Sato} on gemcitabine, a nucleoside anticancer drug.
As such, the proposed GEE methods would provide efficient and robust alternatives for analyzing population PK data in genome-wide PGx studies.

From the simulation results, the type-I error rates of GEE ($\widehat{\mathbf{V}}_s$) were well controlled below the nominal level in all conditions, and were closer to the nominal level than the type-I error rates of GEE ($\widetilde{\mathbf{V}}_s$) despite the downward bias of $\widehat{\mathbf{V}}_s$.
By contrast, in some instances, the type-I error rate of NLMM was inflated;
consequently, GEE ($\widehat{\mathbf{V}}_s$) might be more robust than NLMM under various structures of individual variations.
Therefore, the results of this study show that GEE ($\widehat{\mathbf{V}}_s$) yields valid inference even under inter-individual variability, without assumptions of a random-effects distribution.

In all simulations, GEE ($\widehat{\mathbf{V}}_s$) was computationally fastest and most stable.
In particular, the possible impact of the convergence failures on the type-I error rate and power of NLMM was not clear, and should not be ignored.
Thus, GEE ($\widehat{\mathbf{V}}_s$) is more efficient and computationally stable than NLMM.

In the application to the genome-wide PGx study, GEE ($\widehat{\mathbf{V}}_s$) gave results for all 109,365 SNPs in a relatively short time.
By contrast, a complex NLMM failed to converge and required variable selection.
As the result of additional analyses, many of the significant SNPs detected by GEE ($\widehat{\mathbf{V}}_s$) can also be detected by NLMM.
Therefore, GEE ($\widehat{\mathbf{V}}_s$) can be applied to genome-wide PGx studies, and is remarkably stable and convenient.

The proposed approach may be applicable to other situations.
Because it was formulated based on GEE, it can deal with correlated response data.
Furthermore, it treats inter-individual variability in PK parameters by a potentially misspecified model.
Therefore, it may be applied to correlated response data with inter-individual variability in model parameters.
In particular, when problems occur in association with a strong assumption of random effects, in many cases the proposed approach represents an alternative to mixed models.
However, further research is needed to determine whether this approach is applicable to other settings, because the properties of $\boldsymbol{\beta}_*$ and the performance of the proposed tests are not clear in every particular case.

In summary, this study has demonstrated that GEE ($\widehat{\mathbf{V}}_s$) yields a valid inference even under inter-individual variability, and is more efficient and computationally stable than NLMM.
We conclude that GEE ($\widehat{\mathbf{V}}_s$) represents an alternative approach for analyzing population PK data in genome-wide PGx studies.

\appendix
\section*{\texorpdfstring{APPENDIX A. An evaluation of $\boldsymbol{\beta}_*$}{APPENDIX A. An evaluation of beta*}}
We present a derivation for the properties of a constant $\boldsymbol{\beta}_*$ listed in Section 3.3 Theorem 1.

Using a first-order Taylor expansion of the expectation of equation (2) around $\boldsymbol{\beta}_*=\boldsymbol{\beta}_i$, we get
\[
\mathrm{E}\left[ \sum_{i=1}^K \mathbf{U}_i(\mathbf{Y}_i; \boldsymbol{\beta}_*) \right] \approx
\sum_{i=1}^K \left\{
\mathrm{E}\left[ \mathbf{U}_i(\mathbf{Y}_i; \boldsymbol{\beta}_i) \right] +
\mathrm{E}\left[ \frac{\partial \mathbf{U}_i(\mathbf{Y}_i; \boldsymbol{\beta}_*)}{\partial \boldsymbol{\beta}_*^{\mathrm{T}}} \bigg|_{\boldsymbol{\beta}_*=\boldsymbol{\beta}_i}\right]
(\boldsymbol{\beta}_* - \boldsymbol{\beta}_i)
\right\}.
\]
Here $\mathrm{E}\left[ \mathbf{U}_i(\mathbf{Y}_i; \boldsymbol{\beta}_i) \right]=\mathbf{0}$ and $\mathrm{E}[\partial \mathbf{U}_i(\mathbf{Y}_i; \boldsymbol{\beta}_*) / \partial \boldsymbol{\beta}_*^{\mathrm{T}} \mid_{\boldsymbol{\beta}_*=\boldsymbol{\beta}_i}] = \mathbf{I}_{0i}(\boldsymbol{\beta}_i)$. 
Because by definition 
\[
\mathrm{E}\left[ \mathbf{U}_i(\mathbf{Y}_i; \boldsymbol{\beta}_*) \right] = \mathbf{0},
\]
we arrive at
\[
\sum_{i=1}^K \mathbf{I}_{0i}(\boldsymbol{\beta}_i)(\boldsymbol{\beta}_* - \boldsymbol{\beta}_i)
\approx
\mathrm{E}\left[ \sum_{i=1}^K \mathbf{U}_i(\mathbf{Y}_i; \boldsymbol{\beta}_*) \right]
=
\mathbf{0}.
\]
Hence, 
\[
\boldsymbol{\beta}_* \approx
\left\{ \sum_{i=1}^K \mathbf{I}_{0i}(\boldsymbol{\beta}_i) \right\}^{-1}
\left\{ \sum_{i=1}^K \mathbf{I}_{0i}(\boldsymbol{\beta}_i) \boldsymbol{\beta}_i \right\}.
\]

\section*{\texorpdfstring{APPENDIX B. A asymptotic evaluation of $\widehat{d}$}{APPENDIX B. A asymptotic evaluation of }}
We present a derivation of the d.f. $d$ and the estimator $\widehat{d}$ listed in Section 3.6.

We applied the moment estimator of the d.f. from Fay and Graubard \cite{Fay}.
Assuming $\mathbf{U}^{\mathrm{T}}=$ $(\mathbf{U}_1(\mathbf{Y}_1; \boldsymbol{\beta})$, $\mathbf{U}_2(\mathbf{Y}_2; \boldsymbol{\beta})$, $\ldots$, $\mathbf{U}_K(\mathbf{Y}_K; \boldsymbol{\beta}))^{\mathrm{T}}$ is normally distributed with mean vector $\mathbf{0}$ and covariance matrix $\boldsymbol{\Psi}$, where $\boldsymbol{\Psi}=\mathrm{block\mbox{-}diag}(\boldsymbol{\Psi}_1, \boldsymbol{\Psi}_2, \ldots, \boldsymbol{\Psi}_K)$ is a block-diagonal matrix.
The d.f. can be shown to be $d=\{\mathrm{trace}(\boldsymbol{\Psi}\mathbf{M})\}^2 / \mathrm{trace}(\boldsymbol{\Psi}\mathbf{M}\boldsymbol{\Psi}\mathbf{M})$, where $\mathbf{M}=\mathrm{block\mbox{-}diag}(\mathbf{M}_1, \mathbf{M}_2, \ldots, \mathbf{M}_K)$ is a block-diagonal matrix, and $\mathbf{M}_i$ is defined as follows.
Rewrite $\mathbf{c}^{\mathrm{T}} \widehat{\mathbf{V}}_s \mathbf{c}$ as
\[
\mathbf{c}^{\mathrm{T}} \widehat{\mathbf{V}}_s \mathbf{c}=
K^{-1} \sum_{i=1}^K
\{\mathbf{U}_i(\mathbf{Y}_i; \widehat{\boldsymbol{\beta}})\}^{\mathrm{T}}
\mathbf{M}_i
\{\mathbf{U}_i(\mathbf{Y}_i; \widehat{\boldsymbol{\beta}})\},
\]
where $\mathbf{M}_i=\widehat{\mathbf{D}}_i^{\mathrm{T}} \widehat{\mathbf{V}}_i^{-1} \widehat{\mathbf{D}}_i \mathbf{c} \mathbf{c}^{\mathrm{T}} \widehat{\mathbf{D}}_i \widehat{\mathbf{V}}_i^{-1} \widehat{\mathbf{D}}_i^{\mathrm{T}}$.
Under these assumptions, $d(\mathbf{U}^{\mathrm{T}} \mathbf{M} \mathbf{U}) / \mathbf{c}^{\mathrm{T}} \mathbf{V}_s \mathbf{c}$ is asymptotically distributed as a chi-square random variable with d.f. $d$.
That is, $\mathrm{E}[\mathbf{U}^{\mathrm{T}} \mathbf{M} \mathbf{U}]=\mathbf{c}^{\mathrm{T}} \mathbf{V}_s \mathbf{c}$ and $\mathrm{Var}[d(\mathbf{U}^{\mathrm{T}} \mathbf{M} \mathbf{U}) / \mathbf{c}^{\mathrm{T}} \mathbf{V}_s \mathbf{c}]=d^2 \mathrm{Var}[\mathbf{U}^{\mathrm{T}} \mathbf{M} \mathbf{U}] (\mathbf{c}^{\mathrm{T}} \mathbf{V}_s \mathbf{c})^{-2}=2d$.
Solving these systems of equations, $d=2\{ \mathrm{E}[\mathbf{U}^{\mathrm{T}} \mathbf{M} \mathbf{U}] \}^2 / \mathrm{Var}[\mathbf{U}^{\mathrm{T}} \mathbf{M} \mathbf{U}]$.
Furthermore, $\mathrm{E}[\mathbf{U}^{\mathrm{T}} \mathbf{M} \mathbf{U}]=\mathrm{trace}(\boldsymbol{\Psi}\mathbf{M})$ and $\mathrm{Var}[\mathbf{U}^{\mathrm{T}} \mathbf{M} \mathbf{U}]=2\mathrm{trace}(\boldsymbol{\Psi}\mathbf{M}\boldsymbol{\Psi}\mathbf{M})$ \cite{Searle}, because $\mathbf{U}^{\mathrm{T}} \mathbf{M} \mathbf{U}$ is a quadratic form.
Since $\boldsymbol{\Psi}_i$ can estimate by
\[
\widehat{\boldsymbol{\Psi}}_i=\widehat{\mathbf{D}}_i^{\mathrm{T}} \widehat{\mathbf{V}}_i^{-1} \widehat{\mathbf{S}}_i \widehat{\mathbf{S}}_i^{\mathrm{T}} \widehat{\mathbf{V}}_i^{-1} \widehat{\mathbf{D}}_i,
\]
the estimator is given by $\widehat{d}=\{\mathrm{trace}(\widehat{\boldsymbol{\Psi}}\mathbf{M})\}^2 / \mathrm{trace}(\widehat{\boldsymbol{\Psi}}\mathbf{M}\widehat{\boldsymbol{\Psi}}\mathbf{M})$.

\section*{Acknowledgements}
The authors deeply thank Professor Nan M. Laird at Department of Biostatistics, Harvard School of Public Health for her valuable advice and suggestions.
The authors would also like to thank the Associate Editor and the anonymous reviewers for their helpful comments and suggestions that helped to improve the paper.
The first author thank Professor F. Hashimoto at Faculty of Pharmaceutical Sciences, Josai University for her advice and support.


\begin{thebibliography}{10}
	\providecommand{\url}[1]{\texttt{#1}}
	\providecommand{\urlprefix}{URL }
	\expandafter\ifx\csname urlstyle\endcsname\relax
	\providecommand{\doi}[1]{DOI:\discretionary{}{}{}#1}\else
	\providecommand{\doi}{DOI:\discretionary{}{}{}\begingroup
		\urlstyle{rm}\Url}\fi
	
	\bibitem{Evans2001}
	Evans WE, Johnson JA.
	Pharmacogenomics: the inherited basis for interindividual differences in drug response.
	\emph{Annual Review of Genomics and Human Genetics}  2001; \textbf{2}(1):9--39.
	\doi{10.1146/annurev.genom.2.1.9}.
	
	\bibitem{Evans2003}
	Evans WE, McLeod HL.
	Pharmacogenomics---drug disposition, drug targets, and side effects.
	\emph{New England Journal of Medicine} 2003; \textbf{348}(6):538--549.
	\doi{10.1056/NEJMra020526}.
	
	\bibitem{Innocenti}
	Innocenti F, Undevia SD, Iyer L, Chen PX, Das S, Kocherginsky M, Karrison T, Janisch L, Ram{\'\i}rez J, Rudin CM, \emph{et~al}.
	Genetic variants in the {UDP}-glucuronosyltransferase {1A1} gene predict the risk of severe neutropenia of irinotecan.
	\emph{Journal of Clinical Oncology} 2004; \textbf{22}(8):1382--1388.
	\doi{10.1200/JCO.2004.07.173}.
	
	\bibitem{Wilkinson}
	Wilkinson GR.
	Drug metabolism and variability among patients in drug response.
	\emph{New England Journal of Medicine} 2005; \textbf{352}(21):2211--2221.
	\doi{10.1056/NEJMra032424}.
	
	\bibitem{Wagner}
	Wagner JG.
	\emph{Pharmacokinetics for the pharmaceutical scientist}.
	Lancaster: Technomic Publishing Company, 1993.
	
	\bibitem{Laird}
	Laird NM, Ware JH.
	Random-effects models for longitudinal data.
	\emph{Biometrics} 1982; \textbf{38}(4):963--974.
	
	\bibitem{Lindstrom}
	Lindstrom MJ, Bates DM.
	Nonlinear mixed effects models for repeated measures data.
	\emph{Biometrics} 1990; \textbf{46}(3):673--687.
	
	\bibitem{Davidian1993}
	Davidian M, Gallant AR.
	The nonlinear mixed effects model with a smooth random effects density.
	\emph{Biometrika} 1993; \textbf{80}(3):475--488.
	\doi{10.1093/biomet/80.3.475}.
	
	\bibitem{Davidian1995}
	Davidian M, Giltinan DM.
	\emph{Nonlinear models for repeated measurement data}.
	New York: Chapman \& Hall, 1995.
	
	\bibitem{Vonesh}
	Vonesh EF, Chinchilli VM.
	\emph{Linear and nonlinear models for the analysis of repeated measurements}.
	NewYork: Marcel Dekker, 1996.
	
	\bibitem{Wolfinger}
	Wolfinger RD.
	Fitting nonlinear mixed models with the new {NLMIXED} procedure.
	\emph{Technical {R}eport 287}, SAS Institute, Cary, North Carolina 1999.
	
	\bibitem{Hesselink}
	Hesselink DA, van Gelder T, van Schaik RH, Balk AH, van~der Heiden IP, van Dam T, van~der Werf M, Weimar W, Mathot RA.
	Population pharmacokinetics of cyclosporine in kidney and heart transplant recipients and the influence of ethnicity and genetic polymorphisms in the {MDR-1}, {CYP3A4}, and {CYP3A5} genes.
	\emph{Clinical Pharmacology \& Therapeutics} 2004; \textbf{76}(6):545--556.
	\doi{10.1016/j.clpt.2004.08.022}.
	
	\bibitem{Bosch}
	Bosch TM, Huitema AD, Doodeman VD, Jansen R, Witteveen E, Smit WM, Jansen RL, van Herpen CM, Soesan M, Beijnen JH, \emph{et~al}.
	Pharmacogenetic screening of {CYP3A} and {ABCB1} in relation to population pharmacokinetics of docetaxel.
	\emph{Clinical Cancer Research} 2006; \textbf{12}(19):5786--5793.
	\doi{10.1158/1078-0432.CCR-05-2649}.
	
	\bibitem{Bertrand2009}
	Bertrand J, Comets E, Laffont CM, Chenel M, Mentr{\'e} F.
	Pharmacogenetics and population pharmacokinetics: impact of the design on three tests using the SAEM algorithm.
	\emph{Journal of Pharmacokinetics and Pharmacodynamics} 2009; \textbf{36}(4):317--339.
	\doi{10.1007/s10928-009-9124-x}.
	
	\bibitem{Bertrand2012}
	Bertrand J, Comets E, Chenel M, Mentr{\'e} F.
	Some alternatives to asymptotic tests for the analysis of pharmacogenetic data using nonlinear mixed effects models.
	\emph{Biometrics} 2012; \textbf{68}(1):146--155.
	\doi{10.1111/j.1541-0420.2011.01665.x}.
	
	\bibitem{Pinheiro}
	Pinheiro JC, Bates DM.
	Approximations to the log-likelihood function in the nonlinear mixed-effects model.
	\emph{Journal of Computational and Graphical Statistics} 1995; \textbf{4}(1):12--35.
	\doi{10.1080/10618600.1995.10474663}.
	
	\bibitem{Zhang}
	Zhang H, Lu N, Feng C, Thurston SW, Xia Y, Zhu L, Tu XM.
	On fitting generalized linear mixed-effects models for binary responses using different statistical packages.
	\emph{Statistics in Medicine} 2011; \textbf{30}(20):2562--2572.
	\doi{10.1002/sim.4265}.
	
	\bibitem{Hartford}
	Hartford A, Davidian M.
	Consequences of misspecifying assumptions in nonlinear mixed effects models.
	\emph{Computational Statistics \& Data Analysis} 2000; \textbf{34}(2):139--164.
	\doi{10.1016/S0167-9473(99)00076-6}.
	
	\bibitem{Lesaffre}
	Lesaffre E, Spiessens B.
	On the effect of the number of quadrature points in a logistic random effects model: an example.
	\emph{Journal of the Royal Statistical Society: Series C (Applied Statistics)} 2001; \textbf{50}(3):325--335.
	\doi{10.1111/1467-9876.00237}.
	
	\bibitem{Neuhaus}
	Neuhaus JM, Hauck WW, Kalbfleisch JD.
	The effects of mixture distribution misspecification when fitting mixed-effects logistic models.
	\emph{Biometrika} 1992; \textbf{79}(4):755--762.
	\doi{10.1093/biomet/79.4.755}.
	
	\bibitem{Heagerty}
	Heagerty PJ, Kurland BF.
	Misspecified maximum likelihood estimates and generalised linear mixed models.
	\emph{Biometrika} 2001; \textbf{88}(4):973--985.
	\doi{10.1093/biomet/88.4.973}.
	
	\bibitem{Litiere2007}
	Liti{\`e}re S, Alonso A, Molenberghs G.
	{Type I} and {Type II} error under random-effects misspecification in generalized linear mixed models.
	\emph{Biometrics} 2007; \textbf{63}(4):1038--1044.
	\doi{10.1111/j.1541-0420.2007.00782.x}.
	
	\bibitem{Litiere2008}
	Liti{\`e}re S, Alonso A, Molenberghs G.
	The impact of a misspecified random-effects distribution on the estimation and the performance of inferential procedures in generalized linear mixed models.
	\emph{Statistics in medicine} 2008; \textbf{27}(16):3125--3144.
	\doi{10.1002/sim.3157}.
	
	\bibitem{White}
	White H.
	Maximum likelihood estimation of misspecified models.
	\emph{Econometrica} 1982; \textbf{50}(1):1--25.
	
	\bibitem{Yi}
	Yi GY, Reid N.
	A note on mis-specified estimating functions.
	\emph{Statistica Sinica} 2010; \textbf{20}:1749--1769.
	
	\bibitem{Liang}
	Liang KY, Zeger SL.
	Longitudinal data analysis using generalized linear models.
	\emph{Biometrika}  1986; \textbf{73}(1):13--22.
	\doi{10.1093/biomet/73.1.13}.
	
	\bibitem{Zeger1986}
	Zeger SL, Liang KY.
	Longitudinal data analysis for discrete and continuous outcomes.
	\emph{Biometrics} 1986; \textbf{42}(1):121--130.
	
	\bibitem{Aerts}
	Aerts M, Molenberghs G, Ryan LM, Geys H.
	\emph{Topics in modelling of clustered data}.
	London: Chapman \& Hall/CRC, 2002.
	
	\bibitem{Zeger1988}
	Zeger SL, Liang KY, Albert PS.
	Models for longitudinal data: a generalized estimating
	equation approach.
	\emph{Biometrics} 1988; \textbf{44}(4):1049--1060.
	
	\bibitem{Sato}
	Sato Y, Laird NM, Nagashima K, Kato R, Hamano T, Yafune A, Kaniwa N, Saito Y, Sugiyama E, Kim SR, \emph{et~al}.
	A new statistical screening approach for finding pharmacokinetics-related genes in genome-wide studies.
	\emph{The Pharmacogenomics Journal} 2009; \textbf{9}(2):137--146.
	\doi{10.1038/tpj.2008.17}.
	
	\bibitem{Ziegler}
	Ziegler A, K{\"o}nig IR, Thompson JR.
	Biostatistical aspects of genome-wide association studies.
	\emph{Biometrical Journal} 2008; \textbf{50}(1):8--28.
	\doi{10.1002/bimj.200710398}.
	
	\bibitem{Scheulen}
	Scheulen ME, Hilger RA, Oberhoff C, Casper J, Freund M, Josten KM, Bornh{\"a}user M, Ehninger G, Berdel WE, Baumgart J, \emph{et~al}.
	Clinical phase {I} dose escalation and pharmacokinetic study of high-dose chemotherapy with treosulfan and autologous peripheral blood stem cell transplantation in patients with advanced malignancies.
	\emph{Clinical Cancer Research} 2000; \textbf{6}(11):4209--4216.
	
	\bibitem{DePas}
	De~Pas T, de~Braud F, Danesi R, Sessa C, Catania C, Curigliano G, Fogli S, del Tacca M, Zampino G, Sbanotto A, \emph{et~al}.
	Phase {I} and pharmacologic study of weekly gemcitabine and paclitaxel in chemo-naive patients with advanced non-small-cell lung cancer.
	\emph{Annals of Oncology} 2000; \textbf{11}(7):821--827.
	
	\bibitem{Gabrielsson}
	Gabrielsson J, Weiner D.
	\emph{Pharmacokinetic and pharmacodynamic data analysis: concepts and applications}.
	Stockholm: Swedish Pharmaceutical Press, 2000.
	
	\bibitem{Xu}
	Xu R, O'Quigley J.
	Estimating average regression effect under non-proportional hazards.
	\emph{Biostatistics} 2000; \textbf{1}(4):423--439.
	\doi{10.1093/biostatistics/1.4.423}.
	
	\bibitem{Schempter}
	Schemper M, Wakounig S, Heinze G.
	The estimation of average hazard ratios by weighted {Cox} regression.
	\emph{Statistics in Medicine} 2009; \textbf{28}(19):2473--2489.
	\doi{10.1002/sim.3623}.
	
	\bibitem{Fay}
	Fay MP, Graubard BI.
	Small-sample adjustments for wald-type tests using sandwich estimators.
	\emph{Biometrics} 2001; \textbf{57}(4):1198--1206.
	\doi{10.1111/j.0006-341X.2001.01198.x}.
	
	\bibitem{Fai}
	Fai AHT, Cornelius PL.
	Approximate {$F$}-tests of multiple degree of freedom hypotheses in generalized least squares analyses of unbalanced split-plot experiments.
	\emph{Journal of Statistical Computation and Simulation} 1996; \textbf{54}(4):363--378.
	\doi{10.1080/00949659608811740}.
	
	\bibitem{Schaalje}
	Schaalje GB, McBride JB, Fellingham GW.
	Adequacy of approximations to distributions of test statistics in complex mixed linear models.
	\emph{Journal of Agricultural, Biological, and Environmental Statistics} 2002; \textbf{7}(4):512--524.
	\doi{10.1198/108571102726}.
	
	\bibitem{Mancl}
	Mancl LA, DeRouen TA.
	A covariance estimator for {GEE} with improved small-sample properties.
	\emph{Biometrics} 2001; \textbf{57}(1):126--134.
	\doi{10.1111/j.0006-341X.2001.00126.x}.
	
	\bibitem{MacKinnon}
	MacKinnon JG, White H.
	Some heteroskedasticity-consistent covariance matrix estimators with improved finite sample properties.
	\emph{Journal of Econometrics} 1985; \textbf{29}(3):305--325.
	\doi{10.1016/0304-4076(85)90158-7}.
	
	\bibitem{Chesher}
	Chesher A, Jewitt I.
	The bias of a heteroskedasticity consistent covariance matrix estimator.
	\emph{Econometrica} 1987; \textbf{55}(5):1217--1222.
	
	\bibitem{Kauermann}
	Kauermann G, Carroll RJ.
	A note on the efficiency of sandwich covariance matrix estimation.
	\emph{Journal of the American Statistical Association} 2001; \textbf{96}(456):1387--1396.
	\doi{10.1198/016214501753382309}.
	
	\bibitem{Preisser}
	Preisser JS, Qaqish BF.
	Deletion diagnostics for generalised estimating equations.
	\emph{Biometrika} 1996; \textbf{83}(3):551--562.
	\doi{10.1093/biomet/83.3.551}.
	
	\bibitem{Wakefield}
	Wakefield J, Racine-Poon A.
	An application of {Bayesian} population pharmacokinetic/pharmacodynamic models to dose recommendation.
	\emph{Statistics in Medicine} 1995; \textbf{14}(9):971--986.
	\doi{10.1002/sim.4780140917}.
	
	\bibitem{Sheiner}
	Sheiner LB.
	Analysis of pharmacokinetic data using parametric models. {II}. point estimates of an individual's parameters.
	\emph{Journal of Pharmacokinetics and Pharmacodynamics} 1985; \textbf{13}(5):515--540.
	\doi{10.1007/BF01059333}.
	
	\bibitem{Beal}
	Beal SL, Sheiner LB. Heteroscedastic nonlinear regression.
	\emph{Technometrics} 1988; \textbf{30}(3):327--338.
	\doi{10.1080/00401706.1988.10488406}.
	
	\bibitem{Hirakawa}
	Hirakawa M, Tanaka T, Hashimoto Y, Kuroda M, Takagi T, Nakamura Y. {JSNP}: a database of common gene variations in the {Japanese} population.
	\emph{Nucleic Acids Research} 2002; \textbf{30}(1):158--162.
	\doi{10.1093/nar/30.1.158}.
	
	\bibitem{Searle}
	Searle SR.
	\emph{Matrix algebra useful for statistics}. 
	New York: Wiley, 1982.
	
\end{thebibliography}

\newcommand{\noopsort}[1]{}

\end{document}